\documentclass[aps,twocolumn,amsmath,amssymb]{revtex4-1}
\usepackage{multirow}
\usepackage{graphicx,epsfig,bm}
\usepackage{amsfonts,amsthm,dsfont}
\usepackage{wasysym}

\newcommand{\bo}[1]{\mathbf{#1}}

\newcommand{\co}[1]{\left [ #1 \right ]}
\newcommand{\pa}[1]{\left ( #1 \right )}
\newcommand{\ba}[1]{\left | #1 \right |}

\let \Re \relax
\let \Im \relax
\DeclareMathOperator{\Re}{Re}
\DeclareMathOperator{\Im}{Im}
\DeclareMathOperator{\tr}{Tr}

\begin{document}
\title{Heat asymmetries in nanoscale conductors: The role of decoherence and inelasticity}
\author{Javier Arg\"{u}ello-Luengo}
\affiliation{Institut de F\'{\i}sica Interdisciplin\`aria i Sistemes Complexos
IFISC (CSIC-UIB), E-07122 Palma de Mallorca, Spain}
\author{David S\'anchez}
\affiliation{Institut de F\'{\i}sica Interdisciplin\`aria i Sistemes Complexos
IFISC (CSIC-UIB), E-07122 Palma de Mallorca, Spain}
\author{Rosa L\'opez}
\affiliation{Institut de F\'{\i}sica Interdisciplin\`aria i Sistemes Complexos
IFISC (CSIC-UIB), E-07122 Palma de Mallorca, Spain}

\begin{abstract}
We investigate the heat flow between different terminals in an interacting coherent conductor when inelastic scattering is present.  We illustrate our theory with a two-terminal quantum dot setup. Two types of heat asymmetries are investigated:
electric asymmetry $\Delta_E$, which describes deviations of the heat current in a given contact when
voltages are exchanged, and contact asymmetry $\Delta_C$, which quantifies the difference between the power measured
in two distinct electrodes. In the linear regime, both asymmetries agree and are proportional to the Seebeck coefficient, the latter following at low temperature  a Mott-type formula with a dot transmission renormalized by inelasticity. Interestingly, in the nonlinear regime of transport we find $\Delta_E\neq\Delta_C$ and this asymmetry departure depends on the applied bias configuration. Our results may be important for the recent experiments by Lee \textit{et al.} [Nature (London) \textbf{498}, 209 (2013)], where these asymmetries were measured. 
\end{abstract}

\maketitle

\section{Introduction} 
Thermoelectrical transport at the nanoscale is a phenomenon of wide interest due to its fundamental and applied perspectives~\cite{san14}. From the practical point of view, nanostructures spur a wide range of promising thermoelectric applications such as thermocouples \cite{kim12}, local refrigerators \cite{Sha06}, thermal transistors \cite{Gia06}, and thermal rectifiers \cite{Terr02}, among others. The conversion of waste heat into electricity seems to be more efficient at the nanoscale than at macroscopic scales \cite{her13}.
The fast pursuit toward higher efficiency values of the generated electrical power in relation to the supplied heat has reached remarkable results \cite{Jeff07,npg}. However, related fundamental issues such as the electronic heat flow traversing a nanodevice still remain poorly understood mainly because thermal current is not easily accessible in an experiment~\cite{jez13}. In many aspects, heat flow inherently differs from its electrical counterpart and can reveal information about the number of channels available for transport~\cite{mol92}, the presence of interactions~\cite{kan97},
properties of single-particle wave functions~\cite{bat13}, or even superconducting phase differences~\cite{spi14}.

Recent works have investigated both experimentally and theoretically the heat current in atomic-scale junctions~\cite{Lee13,zot14}. Importantly, power dissipation at atomic scales depends strongly on the way in which the transmission probability varies with energy. Thus, for nanostructures showing a strongly energy-dependent transmission, the measured heat flux is shared quite asymmetrically among the contacts whereas for those systems with a weakly energy-dependent transmission the heat asymmetry is strongly suppressed~\cite{Lee13}.  This conclusion assumes that energy exchange between carriers occurs elastically. However, in molecular or atomic junctions, inelastic processes can be of critical importance when internal degrees of freedom such as rotational or vibrational modes come into play~\cite{pau03,fre04,koc04,gal07,fin09,Entin12,zim14}. As a consequence, these can alter the physical scenario. The fundamental question addressed in this work is precisely how the heat current asymmetry is affected by inelastic processes. We are interested in the contact asymmetry $\Delta_C$, which measures differences between the source ($\mathcal{J}_1$) and the drain ($\mathcal{J}_2$) heat currents,
and the electric asymmetry $\Delta_E$, which quantifies the heat-current asymmetry in a given electrode when the applied voltages $V_1$ and $V_2$ are exchanged:
\begin{align}
\Delta_C &=\mathcal{J}_1(V_1,V_2)-\mathcal{J}_2(V_1,V_2)\label{eq_Deltac}\,, \\
 \Delta_E &=\mathcal{J}_1(V_1,V_2)-\mathcal{J}_1(V_2,V_1)\label{eq_Deltae}\,.
\end{align}
Furthermore, even when only elastic processes are present, dephasing mechanisms can also take place. Therefore, it is also natural to ask how heat is partitioned among the different electronic reservoirs in the presence of dephasing. 

To examine both issues, we use the voltage \cite{Butt88,ama90} and dephasing \cite{jon96,lan97} probe models, recently generalized to treat heat-current flows~\cite{Saito11,dav11,Caso12,bed13,ber13,ape13,bra13,Meair14}. In these formulations, inelastic and dephasing processes are incorporated by considering a fictitious terminal attached to the quantum system in such a way that the net electrical and heat currents flowing through the probe vanish. In particular, for a voltage probe a carrier that enters the probe with a given energy is reemitted into the conductor with an unrelated energy. In contrast, when only dephasing processes are present, the energy-resolved heat and charge currents are identically zero at each energy.
Since the model is independent of the microscopic details of the actual scattering mechanisms, the results are simple
to understand and can be applied to a large variety of systems.
\begin{figure}
\begin{center}
\includegraphics[width=0.4\textwidth]{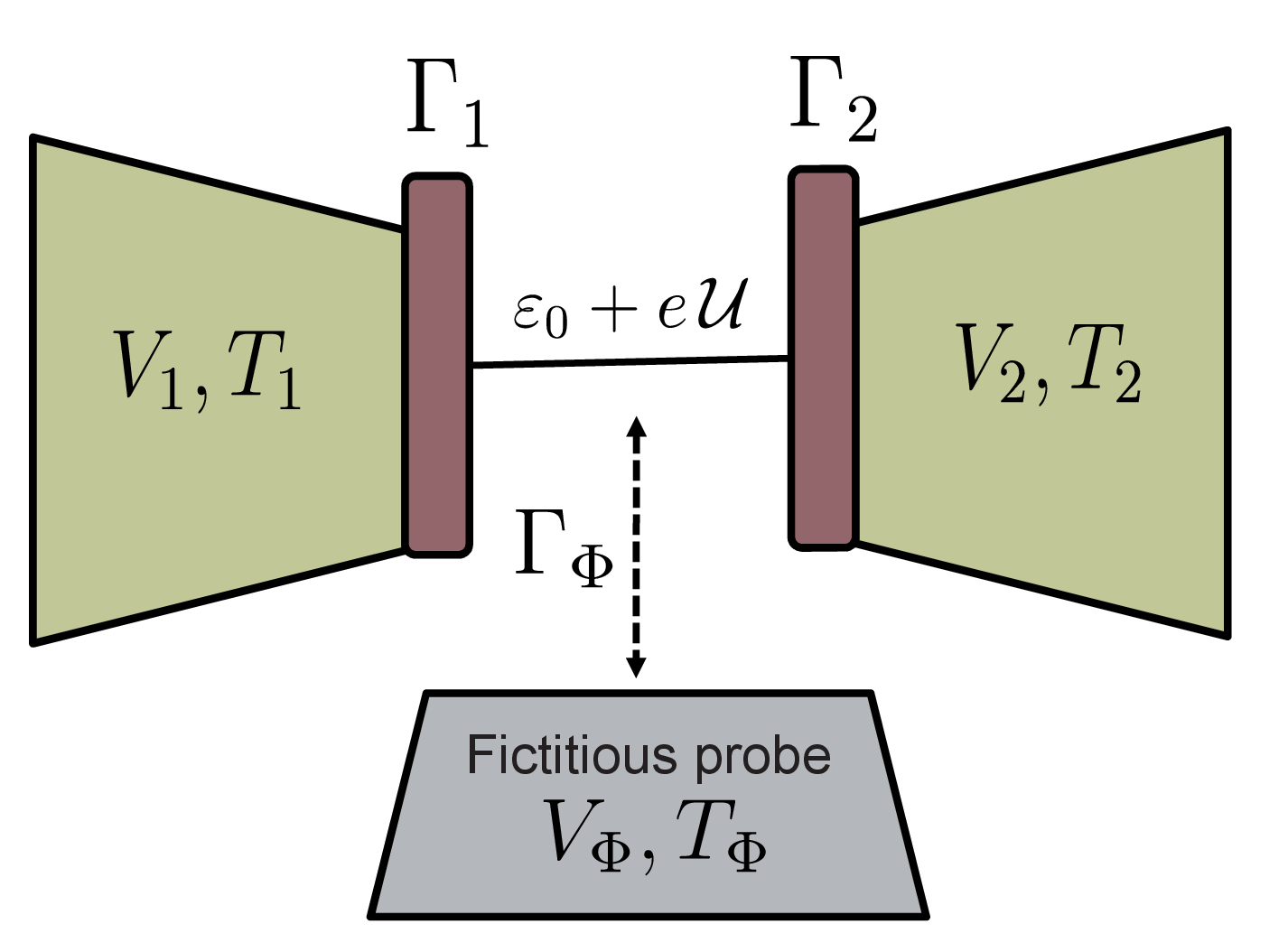}
\caption{Schematic of a generic nanoconductor with energy level $\varepsilon_0$
in the presence of a voltage $V_\Phi$ and temperature $T_\Phi$
probes and coupled to left and right reservoirs by tunnel couplings $\Gamma_1$ and $\Gamma_2$. Here, $\Gamma_{\Phi}$ is the tunnel coupling with the probe. $V_\Phi$ and $T_{\Phi}$ adjust themselves in order to
cancel the net flow of heat and charge through the probe. The internal potential in the conductor is denoted with ${\cal U}$.}
\label{Figure1}
\end{center} 
\end{figure}

Our theory is illustrated with a prototypical model for mesoscopic systems: a localized state (representing many different quantum systems, i.e., atomic or molecular junctions, quantum dots, etc.) attached to two electronic reservoirs and subject to different chemical and temperature biases, as depicted in Fig.~\ref{Figure1}. 

\section{Theoretical model} 
When a mesoscopic conductor is coupled to $i=1\cdots N$ electronic reservoirs and is driven out of equilibrium by electrostatic fields $\{V_{i}\}$ or temperature gradients $\{\theta_i\}$, a flow of charge and energy from  the reservoirs  toward the conductor is established. Charge conservation dictates that all stationary charge flows add up to zero, $\sum_{i=1\cdots N} \mathcal{I}_{i}=0$, whereas the sum of thermal currents must include the Joule heating term, $\sum_{i=1\cdots N} (\mathcal{J}_{i}+\mathcal{I}_{i}V_{i})=0$.  Within the scattering approach formalism, the flows read as~\cite{butcher}
\begin{eqnarray}\label{currentheat}
&&\mathcal{I}_{i}=\frac{2e}{h}\sum_{j} \int dE A_{ij} f_{j}(E)\,,
\\ 
&&\mathcal{J}_{i}=\frac{2}{h} \sum_{j} \int dE \left(E-\mu_i\right) A_{ij} f_{j}(E)\,.
\label{flows}
\end{eqnarray}
The factor $2$ originates from spin degeneracy since we do not consider external magnetic fields.
The heat current $\mathcal{J}_{i}=\mathcal{J}_{i}^{E}-V_{i}\mathcal{I}_{i}$ is given by sum of the energy current $\mathcal{J}_{j}^{E}=(2/h) \sum_{j} \int dE (E-E_F) A_{ij} f_{j}(E)$ and the associated Joule dissipating heat power $V_{j}\mathcal{I}_j$. The electrochemical 
potential in reservoir $i$ is defined as $\mu_i=E_F+e V_{i}$ with $E_F$ the Fermi energy,
and $f_j(E)=\left[1+\exp{\pa{(E-\mu_j)/k_B T_j}}\right]^{-1}$ is the Fermi-Dirac distribution function.  
Each terminal has temperature $T_i=\theta+\theta_i$, obtained from a temperature shift $\theta_i$ with respect to the background temperature $\theta$. The elements  $A_{ij}=\tr [\delta_{ij}- s_{ij}^\dagger(E,\mathcal{U}(\vec{r}, \{V_{k}\},\{\theta_k\})s_{ij}(E,\mathcal{U}(\vec{r}, \{V_{k}\},\{\theta_k\})]$  (with $k=1\cdots N$)
are given in terms of  the scattering matrix $s$, where
$\tr s^\dagger_{ij}s_{ij}=T_{ij}$ is the transmission probability from terminal
$j$ to contact $i$ and the trace is performed over the contact channels. Importantly, due to electronic repulsion the potential profile inside the conductor is altered when charge is injected by means of electrical or thermal biases. As a result, the scattering properties of the conductor expressed by $s_{ij}(E,e\mathcal{U} (\vec{r}, \{V_{k}\},\{\theta_k\})$  depend not only on the carrier energy $E$ but also on the internal potential landscape $\mathcal{U}$, which depends in turn on the set of voltage and temperature shifts. The electrostatic response can be determined from the Poisson equation $\varepsilon_{v}\nabla^2 \mathcal{U}(\vec{r})=-q$ with $\varepsilon_{v}$ the vacuum permittivity and $q$ the total charge inside the conductor built up from (bare) charges injected by electrical and thermal gradients and screened charges created in response to the external perturbations~\cite{chr96,san13,jac13}.

At sufficiently low biases in the applied voltages and temperatures, Eq.~\eqref{currentheat} is expanded
up to first order in the shifts $V_i$ and $\theta_i$ and the result can be expressed in matrix form:
\begin{equation}\label{eq_matrix}
\pa{
\begin{array}{c}
\bm{\mathcal{I}} \\ 
\bm{\mathcal{J}}
\end{array} }
=\pa{
\begin{array}{cc}
\bo{G} & \bo{L} \\ 
\bo{M} & \bo{K}
\end{array}}
\pa{
\begin{array}{c}
\bo{V} \\
\bm{\theta}
\end{array} }\,,
\end{equation}
where we have defined the vectors $\bm{\mathcal{I}}=\co{\mathcal{I}_1,\cdots,\mathcal{I}_N}^T$, $\bm{\mathcal{J}}=\co{\mathcal{J}_1,\cdots,\mathcal{J}_N}^T$, $\bo{V}=\co{V_1,\cdots,V_N}^T$ and $\bm{\theta}=\co{\theta_1,\cdots,\theta_N}^T$. The elements of the submatrices $\bo{G}$, $\bo{L}$, $\bo{M}$ and $\bo{K}$
are the transport coefficients
\begin{align}
G_{ij} &= \frac{2e^2}{h} \int\! dE  \,(\mathcal{N}_i \,\delta_{ij}-T_{ij}) (-f_{\rm eq}')\label{eq_rc1}\,,\\
L_{ij} &= \frac{2e}{h\theta} \int\!dE \,(E-E_F)(\mathcal{N}_i\,\delta_{ij}-T_{ij})(-f_{\rm eq}')\label{eq_rc2}\,,\\
M_{ij}&=\theta L_{ij}\,,\label{eq_rc3}\\
K_{ij} &= \frac{2}{h\theta} \int\!dE\, (E-E_F)^2 (\mathcal{N}_i\,\delta_{ij}-T_{ij})(-f_{\rm eq}')\label{eq_rc4}\,,
\end{align}
where $\mathcal{N}_i$ represents the channel number in the $i$-th contact and $f_{\rm eq}'$ denotes
the energy derivative of the Fermi distribution function evaluated at $V_i=\theta_i=0$.
Equation~\eqref{eq_rc3} is a consequence of reciprocity. Additionally, the transmission in
the linear-response regime is evaluated at the equilibrium potential and is thus independent
of the nonequilibrium screening $\mathcal{U}$. We will later consider the nonlinear regime,
in which currents do depend on $\mathcal{U}$.

\section{Elastic and inelastic probes} To include inelastic processes in the thermoelectric transport we consider an additional fictitious probe, denoted by $\Phi$, that plays simultaneously the role of an ideal voltmeter and thermometer. Then, both charge $\mathcal{I}_{\Phi}$ and heat $\mathcal{J}_{\Phi}$ currents through the probe are identically zero. Each current carrier absorbed into the probe is reemitted with unrelated phase and energy. We hence use the conditions $\mathcal{I}_\Phi=\mathcal{J}_{\Phi}=0$ to eliminate
the probe voltage $V_\Phi$ and temperature $T_\Phi$ and rewrite Eq.~\eqref{eq_matrix}
with modified transport coefficients:
\begin{align}
\tilde{G}_{ij}&=G_{ij}+D [G_{i\Phi}(L_{\Phi\Phi}M_{\Phi j}-K_{\Phi\Phi} G_{\Phi j}) \nonumber\\
&+L_{i\Phi}(M_{\Phi\Phi}G_{\Phi j}-G_{\Phi\Phi} M_{\Phi j})]\,,\\
\tilde{L}_{ij}&=L_{ij}+D [G_{i\Phi}(L_{\Phi\Phi}K_{\Phi j}-K_{\Phi\Phi} L_{\Phi j}) \nonumber\\
&+L_{i\Phi}(M_{\Phi\Phi}L_{\Phi j}-G_{\Phi\Phi} K_{\Phi j})]\,,\\
\tilde{K}_{ij}&=K_{ij}+D [K_{i\Phi}(M_{\Phi\Phi}L_{\Phi j}-G_{\Phi\Phi} K_{\Phi j}) \nonumber\\
&+M_{i\Phi}(L_{\Phi\Phi}K_{\Phi j}-K_{\Phi\Phi} L_{\Phi j})]\,,
\end{align}
and $\tilde{M}_{ij}=\theta \tilde{L}_{ij}$ insofar as the Kelvin-Onsager symmetry condition is preserved even in the presence of the probe. Here, $D=(G_{\Phi\Phi}K_{\Phi\Phi}-L_{\Phi\Phi} M_{\Phi\Phi})^{-1}$.

When the source of scattering is elastic, one employs a dephasing probe. The charge- (heat-) current density $i(E)$ [$j(E)$]
is determined from the equation $\mathcal{I}(E)=\int dE\, i(E)$ [$\mathcal{J}(E)=\int dE\, j(E)$]. We impose the condition that for each energy $E$ the probe
draws no net current, $i_\Phi (E)=j_\Phi (E)=0$, resulting in the unique probe distribution function
$f_{\Phi}=-\sum_i A_{\Phi i}f_i/A_{\Phi \Phi}$. Substituting $f_{\Phi}$ back into the charge and heat flows, one arrives at
\begin{eqnarray}\label{chargeheatelastic2}
\mathcal{I}_{i}&=&\frac{2e}{h}\sum_{j} \int dE \pa{A_{ij}-\frac{A_{i\Phi}A_{\Phi j}}{A_{\Phi\Phi}}} f_j \,,
\\ 
\mathcal{J}_{i}&=&\frac{2}{h}\sum_{j} \int dE \pa{E-\mu_i} \pa{A_{ij}-\frac{A_{i\Phi}A_{\Phi j}}{A_{\Phi\Phi}}}f_j \,.
\label{chargeheatelastic}
 \end{eqnarray}
Here, $A_{ij}-A_{i\Phi}A_{\Phi j}/A_{\Phi\Phi}$ includes a transmission function
renormalized by decoherence effects due to the probe coupling. 

\section{Source-drain conductors} In the following, we focus on a simple geometry: a two-terminal conducting device as illustrated in Fig.~\ref{Figure1}. Let $V_{1}$ ($V_2$) be the bias drop and temperature applied to terminal $1$ ($2$) in the isothermal case
($\theta_{1}=\theta_2=\theta$).
The power measured at each contact is shown to exhibit different values depending on the configuration measurement~\cite{Lee13}.  
We commence our analysis with the linear regime in which voltage shifts are very small. Due to energy current conservation, the condition $\mathcal{J}_1(V_1,V_2)=-\mathcal{J}_2(V_1,V_2)$ holds. Hence, the measured heat contact asymmetry [Eq.~\eqref{eq_Deltac}] in the absence of incoherent scattering becomes
\begin{eqnarray}\label{eq_DeltaC}
\Delta_C =2M_{11}V=-2S G_{11}V\theta,
\end{eqnarray}
to leading order in $V=V_1-V_2$. Corrections would be of the order of $V^2$. In Eq.~\eqref{eq_DeltaC}, $S=-L_{11}/G_{11}$ represents the Seebeck coefficient. The asymmetry is proportional to the thermopower~\cite{Lee13} since $S$ indeed measures the asymmetry between electron-like and hole-like transport. Importantly, the contact asymmetry [Eq.~\eqref{eq_Deltac}] amounts to the electrical asymmetry [Eq.~\eqref{eq_Deltae}] due precisely to the energy conservation condition. Now, in the presence of inelasticity (voltage probe) we find that both heat asymmetries still coincide ($\Delta\equiv\Delta_C=\Delta_E$) and are given by
\begin{align}
\Delta &= 2 L_{11}V\theta +2 D V \theta [G_{\Phi 1} L_{\Phi \Phi} K_{1\Phi}-
G_{\Phi 1} K_{\Phi \Phi} L_{1\Phi}
\nonumber\\
&- L_{\Phi 1} G_{\Phi \Phi} K_{1\Phi}+\theta L_{1\Phi} L_{\Phi \Phi}L_{\Phi 1}]\,,
\label{asim}
\end{align}
which is valid up to linear order in $V$.
Clearly, when the probe is decoupled we recover Eq.~\eqref{eq_DeltaC}.

At low temperature, a Sommerfeld expansion of Eq.~\eqref{asim} yields
\begin{equation}\label{asyminelastic}
\Delta=\frac{4\pi^2 e k_B^2 V\theta^2}{3h} \left (T_{12}+\frac{T_{1\Phi}T_{\Phi 2}}{T_{1\Phi}+T_{2\Phi}}\right)'+\mathcal{O}(\theta^4)\,,
\end{equation}
where the prime indicates that the energy derivative is evaluated at $E=E_F$.
Equation~(\ref{asyminelastic}) has a surprisingly simple form. We recall that in the presence of a voltage probe the transmission
is split into the coherent term $T_{12}$ associated with those carriers that flow between source and drain without interacting with the probe
and the incoherent transmission $T_{1\Phi}T_{\Phi 2} / (T_{1\Phi}+T_{2\Phi})$, which accounts for the fraction of carriers that are incoherently scattered through the probe $\Phi$~\cite{Butt88}.
Here, we find that the heat asymmetry is nicely given by the energy derivative of both terms summed.
In fact, Eq.~\eqref{asyminelastic} can be interpreted as a Mott-type formula in which $S$, which is proportional to $T_{12}'$
in Eq.~\eqref{eq_DeltaC} due to the Mott relation~\cite{mott,*jon80}, becomes modified by the incoherent part but keeping the same structural form.

For the dephasing probe, we first perform a linear expansion for $V_{ij}$ in Eq.~\eqref{chargeheatelastic}. Then, the heat asymmetry reads as
\begin{equation}\label{asyelastic}
\Delta =\frac{4eV}{h} \int dE \,(E-E_F) \left (T_{12}+\frac{T_{1\Phi}T_{\Phi 2}}{T_{1\Phi}+T_{2\Phi}}\right) (-f_{\rm eq}')\,.
\end{equation}
An important remark here is in order. The heat asymmetry for the inelastic probe and the dephasing case differ at temperatures higher than the energy scale at which the renormalized transmission varies appreciably. However, to lowest order in the background temperature, Eq.~(\ref{asyelastic}) identically gives Eq.~(\ref{asyminelastic}). This implies that at low temperature, the heat asymmetry is largest when the renormalized transmission (i.e., the coherent plus the incoherent terms) varies rapidly with energy around $E_F$
and that both dephasing and inelastic mechanisms contribute equally. Deviations appear to higher order in $\theta$.
Importantly,  when all transmissions are functions of the local density of states~\cite{mei91}, the asymmetry $\Delta$
cancels out in the electron-hole symmetry case.

\section{Nonlinear heat asymmetries: A quantum dot example}
The nonlinear regime of thermoelectric transport shows unique effects~\cite{san13,jac13,Star93,Sve13,sie14}.
For the heat transport, rectifications have attracted a good deal of attention~\cite{kul94,bog99,li04,seg06,ruo09,kuo10,whi13,lop13,hwa13,whi14,uts14,wer14,eic14,ger14,bie14}.
Crucially, electron-electron interactions must now be taken into account.
It is worthy to mention that a description of the contact heat asymmetry in terms of a probe-renormalized transmission is then no longer possible. Instead, we need to self-consistently find the internal potential of the conductor. For this purpose, we illustrate the heat asymmetries for the relevant case of a quantum dot coupled to two reservoirs. A quantum dot model is the basic description of atomic and molecular junctions in terms of localized atomic or molecular orbitals~\cite{cue10}. The scattering matrix is modeled as a Breit-Wigner resonance,
\begin{equation}\label{eq_s}
s_{ij}(E)=\delta_{ij}-\frac{i\sqrt{\Gamma_{i}\Gamma_{j}}}{E-\varepsilon_0+i\Gamma/2}\,,
\end{equation}
centered at the atomic/molecular orbital level position $\varepsilon_0$. Here, $\Gamma_{i}$ denotes the tunneling rates when the localized level is coupled to the left and right reservoirs ($\Gamma_i=\Gamma_1,\Gamma_2$) and $\Gamma_0=\Gamma_1+\Gamma_2$. The total broadening is thus $\Gamma=\Gamma_0+\Gamma_\Phi$, where the dot coupling to the fictitious probe $\Gamma_\Phi$ quantifies the degree of inelastic/dephasing processes in our transport description. Due to the simplicity of the Breit-Wigner model, the results for the dephasing and voltage/temperature coincide. We leave open the question of having different heat asymmetry responses for more intricate setups.

The internal potential $\mathcal{U}$ is assumed to be spatially homogeneous. We thus make the substitution $\varepsilon_0\to\varepsilon_0+e\mathcal{U}$ in Eq.~\eqref{eq_s}. $\mathcal{U}$ is determined from a discretized version of the Poisson equation in terms of a capacitance $C$: $C\mathcal{U}=q_d-q_{\rm eq}$, where $q_d$ is the nonequilibrium dot charge
\begin{equation}\label{eq_charge}
q_d=\frac{2e}{\pi}\int dE \frac{\Gamma_1f_1+\Gamma_2f_2+\Gamma_\Phi f_\Phi}{(E-\varepsilon_0-e{\cal U})^2+\Gamma^2/4}\,\,,
\end{equation}
and $q_{\rm eq}$ follows from Eq.~\eqref{eq_charge} by setting all voltages and temperature shifts to zero ($f_1=f_2=f_{\rm eq}$
and thereby $f_\Phi=f_{\rm eq}$). Note that
$q_d$ is a nonlinear function of the thermoelectric configuration and that $U$ depends implicitly on voltage and thermal biases.

To compute the heat flow in the presence of interactions and inelastic processes, a system of three nonlinear equations are to be solved simultaneously: the capacitance equation to obtain $\mathcal{U}(\{V_k\}, \{\theta_{k}\}))$ and the two conditions for the fictitious probe, $\mathcal{I}_\Phi=0$ and $\mathcal{J}_{\Phi}=0$, that determine $V_{\Phi}$ and $T_{\Phi}$. Then, $V_{\Phi}(\{{V}_{i},{\theta}_{i}\})$ and $T_{\Phi}(\{{V}_{i},{\theta}_{i}\})$ are nonlinear functions of the shifts applied to the electrodes. Once these parameters are self-consistently obtained, heat-flow asymmetries can be investigated. Remarkably and in contrast to the linear regime, the contact $\Delta_{C}$ and electrical $\Delta_{E}$ asymmetries do not generally coincide,
\begin{align}\label{asymmetries}
\Delta_{C}&=2 \mathcal{J}^{E}(V_1,V_2)-(V_1+V_2) \mathcal{I}(V_1,V_2) \,, \\
\Delta_{E}&=\mathcal{J}^{E}(V_1,V_2)-\mathcal{J}^{E}(V_2,V_1)\nonumber \\
&-V_1 \mathcal{I}(V_1,V_2)+V_2 \mathcal{I}(V_2,V_1)\,.
\end{align}
Here, we define $\mathcal{J}^{E}(V_1,V_2)=\mathcal{J}_{1}^E(V_1,V_2)=-\mathcal{J}_{2}^E(V_1,V_2)$, and $\mathcal{I}=\mathcal{I}_1=-\mathcal{I}_2$. Note that $\Delta_C$ depends on the particular way in which electrical biases are applied. For a symmetric electrical bias configuration, $\Delta_C$ is indeed a measure of the energy current. Both asymmetries agree as long as the transmission is symmetric under the transformation $V_1\leftrightarrows V_2$, which leads to
odd charge currents under reversing the bias polarity. However, this condition is not generally met when interactions are present and rectification effects then arise. Importantly, the Joule heating term affects differently the two asymmetries and is, in many cases, the dominant contribution, as we demonstrate in the following.
\begin{figure}
\begin{center}
\hspace*{-0.5cm} 
\includegraphics[width=0.42\textwidth, angle =-90]{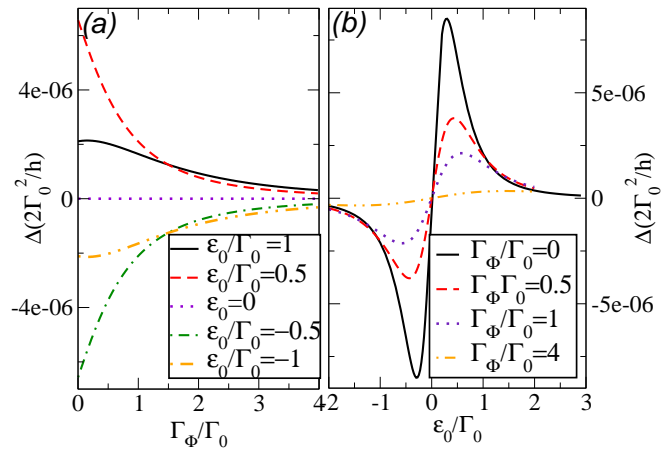}
\caption{Heat-current asymmetry  in the linear regime (a) as a function of the coupling of the probe $\Gamma_\Phi/\Gamma_0$, and (b)  versus the localized level $\varepsilon_0/\Gamma_0$. Parameters: $E_F=0$, $\Gamma_1=\Gamma_2=\Gamma_0/2$, $eV=0.01\Gamma_0$, $k_B \theta=0.01\Gamma_0$.}
\label{Figure2}
\end{center}
\end{figure}

\section{Numerical results}

In this section, we present numerical calculations for the heat asymmetries of our two-terminal quantum dot
in both linear and nonlinear regimes.
In either case, the integrals following from substitution of Eq.~\eqref{eq_s} in Eqs.~\eqref{currentheat} and~\eqref{flows}
require a careful analysis (see Appendix). We begin with the linear regime. 
In Fig.~\ref{Figure2}, we show the heat asymmetry  $\Delta=\Delta_C=\Delta_E$ as a function of the probe coupling $\Gamma_\Phi/\Gamma_0$ [Fig.~\ref{Figure2}(a)] and the dot level $\varepsilon_0/\Gamma_0$, which can be tuned with an external gate voltage [Fig.~\ref{Figure2}(b)].
We observe in Fig. \ref{Figure2}(a) that inelastic processes reduce the heat asymmetry $\Delta$ and that the asymmetry is not a monotonic function of the gate. Due to the probe coupling, the dot transmission acquires an additional level broadening (we recall that $\Gamma=\Gamma_1+\Gamma_2+\Gamma_\Phi$). When $\Gamma_\Phi$ increases the dot transmission becomes broader and shows a weaker energy dependence. As a result, the energy current is reduced overall. We notice that $\Delta$ shows electron-hole symmetry, i.e., $\Delta(\varepsilon_0)=-\Delta(-\varepsilon_0)$.  This fact is more explicit in Fig.~\ref{Figure2}(b) and is due to the absence of screening effects in the linear regime. Here, the curves $\Delta$ versus the dot level show a resonant-like behavior in which $\Delta$ becomes an extremum for $2|\varepsilon_0|/\Gamma\simeq 1$.  Additionally, this $\Delta$ extremal point depends quite strongly on $k_B\theta$ (not shown here). Indeed, at very low temperatures the value for which $\Delta$ is maximum or minimum indicates the energy scale  for which the transmission changes more abruptly around the Fermi energy.
\begin{figure}
\begin{center}
\hspace*{-0.3cm} 
\includegraphics[width=0.4\textwidth, angle=-90]{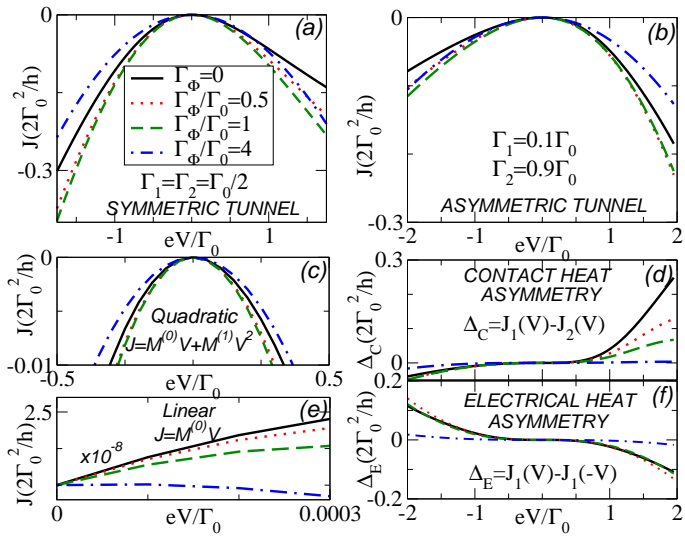}
\caption{Heat-current characteristic  ($\mathcal{J}-V$) for various values of the probe strength $\Gamma_\Phi/\Gamma_0$ (a) for a symmetric tunnel configuration $\Gamma_1=\Gamma_2=\Gamma_0/2$, and (b) for an  asymmetric tunnel configuration $\Gamma_1=0.1\Gamma_0$, and $\Gamma_2=0.9\Gamma_0$. (c), and (e) show enlarged parts of  $\mathcal{J}-V$ corresponding to the quadratic and linear behavior of the heat current versus voltage, respectively. (d), and (f) illustrate the contact ($\Delta_C$), and electrical ($\Delta_E$) heat asymmetries.  Parameters: $E_F=0$, $k_B\theta=0.01\Gamma_0$, $\varepsilon_0/\Gamma_0=1$, $C=0$. }
\label{Figure3}
\end{center}
\end{figure}

In the nonlinear regime, rectification effects arise, as illustrated in Fig.~\ref{Figure3}. For definiteness, we consider a symmetrically electrical biased quantum dot, i.e., $V_1=-V_2=V/2$ with a common background temperature  $\theta$ for both contacts. This electrical and thermal configuration mimics the experimental conditions reported by Lee  {\it et al.}  in Ref.~\cite{Lee13}.  We show the heat flow $\mathcal{J}=\mathcal{J}_1(V)$ through contact $1$ for a symmetrically coupled quantum dot  ($\Gamma_1=\Gamma_2$) in Fig.~\ref{Figure3}(a), and for an asymmetric tunnel configuration in Fig.~\ref{Figure3}(b) ($\Gamma_1\neq\Gamma_2$). In both cases, we observe rectification effects, $\mathcal{J}(V)\neq -\mathcal{J}(-V)$ even for moderate voltages. These are mainly caused by the Joule heating term, which can be further strengthened by an asymmetric potential response in the case $\Gamma_1\neq\Gamma_2$~\citep{lop13}. In fact, as shown in Fig. \ref{Figure3}(c), the heat flow becomes a quadratic function of voltage, $\mathcal{J}(V)=M^{(0)}V+ M^{(1)}V^2$ ($M^{(0)}=M_{11}$ represents the leading-order electrothermal coefficient~\cite{lop13}). Thus, $J$ is quickly dominated by the Joule power at low bias $\mathcal{P}_{\rm Joule}=\mathcal{I}V\propto M^{(1)}V^2$. For a small-bias range, Fig.~\ref{Figure3}(e) displays the linear transport regime in which $\mathcal{J}=M^{(0)}V\propto V$ (Peltier effect). We observe that in the strongly nonlinear regime, the effect of increasing $\Gamma_\Phi/\Gamma_0$ [Figs.~\ref{Figure3}(a) and~\ref{Figure3}(b)] causes a decrease of the Peltier and the heat current thus becomes more symmetric under reversal of the bias polarity. 
The contact and electric heat asymmetries, $\Delta_C$ and $\Delta_E$, are shown in Figs.~\ref{Figure3}(d) and~(f). We observe that inelastic processes (increasing $\Gamma_\Phi/\Gamma_0$) reduce the value of $\Delta_C$ [Fig.~\ref{Figure3}(d)] since the contact heat asymmetry $\Delta_C$ coincides with the energy current $\mathcal{J}^{E}$ for symmetric biases, as shown in Eq.~(\ref{asymmetries}). Hence, by increasing $\Gamma_{\Phi}/\Gamma_0$  the transmission acquires a weaker energy dependence, leading to a suppression of the energy current for our device and therefore a decrease of $\Delta_C$. The electrical heat asymmetry is, by construction, insensitive to rectification effects, as depicted in Fig.~\ref{Figure3}(f). Moreover, we also observe a decrease of $\Delta_E$ as the amount of incoherent scattering, $\Gamma_{\Phi}/\Gamma_0$, increases.
\begin{figure}
\begin{center}
\hspace*{-0.2cm} 
\includegraphics[width=0.42\textwidth, angle =-90]{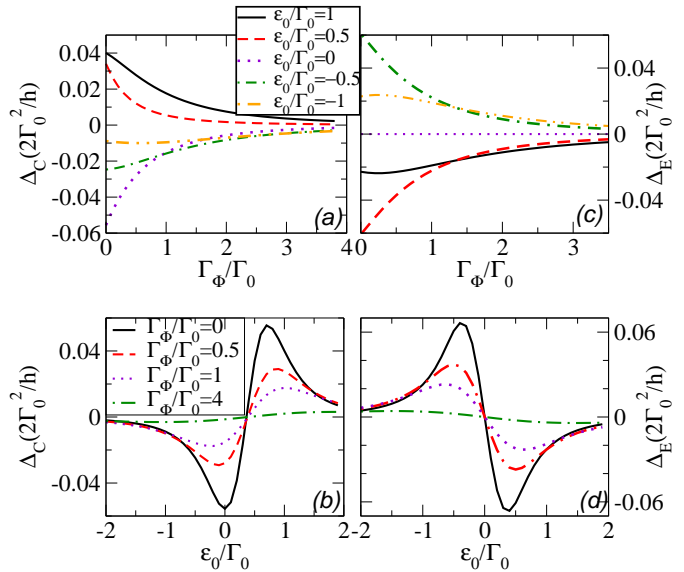}
\caption{Heat-current asymmetry in the nonlinear regime for an asymmetric tunnel configuration $\Gamma_1=0.1\Gamma_0$, and $\Gamma_2=0.9\Gamma_0$. Contact heat asymmetry $\Delta_C=J_1(V)-J_2( V)$: (a) as a function of the coupling of the probe $\Gamma_\Phi/\Gamma_0$, and (b)  versus the localized level $\varepsilon_0/\Gamma_0$. Electrical heat asymmetry
$\Delta_E=J_1(V)-J_1( -V)$: (c) as a function of the probe coupling $\Gamma_\Phi/\Gamma_0$, and (d) versus the dot level $\varepsilon_0/\Gamma_0$. Parameters: $E_F=0$, $\Gamma_1=0.1\Gamma_0$, $\Gamma_2=0.9\Gamma_0$, $eV=\Gamma_0$, $k_B\theta=0.01\Gamma_0$, $C=0$.}
\label{Figure4}
\end{center}
\end{figure}

Finally, we discuss the behavior of the heat asymmetries with the probe coupling strength $\Gamma_\Phi$
and dot level $\varepsilon_0$ in Fig.~\ref{Figure4} for the nonlinear regime (we set $eV/\Gamma_0=1$). The heat-contact asymmetry 
dependence with $\Gamma_\Phi/\Gamma_0$ is presented in Fig.~\ref{Figure4}(a) for different 
$\varepsilon_0/\Gamma_0$ values. We observe departures from the electron-hole symmetry,
$\Delta_C(\varepsilon_0)\neq -\Delta_C(-\varepsilon_0)$, when the strength of the probe is relatively small, whereas if $\Gamma_{\Phi}/\Gamma_0$ becomes larger such effects are removed due to an overall reduction of 
this heat asymmetry. Figure~\ref{Figure4}(c) shows the electrical asymmetry, which is electron-hole symmetric by 
construction. As previously, $\Delta_E$ is broadly reduced when $\Gamma_\Phi/\Gamma_0$ grows. The dot gate 
dependence of the heat asymmetries, for specific values of $\Gamma_\Phi/\Gamma_0$, is shown in Figs.~\ref{Figure4}(b) and~(d).
In both cases, the heat asymmetries show a peak structure, which is 
reduced with increasing probe strengths. However, Fig.~\ref{Figure4}(b) clearly shows the absence of electron-hole symmetry
for $\Delta_C$.

\section{Conclusions}
 In closing, we have formulated a generic framework for the assessment of inelastic and dephasing processes in
 the power asymmetry of nanoscale junctions. We have found that in linear response,
 the heat-current asymmetries (both measured in a given contact or in different electrodes) agree and
 are given at low temperatures by the energy derivative of a modified transmission function.
 In the nonlinear regime of transport, both asymmetries differ and present an interesting behavior
 in terms of the coupling to the dephasing probe and the gate-tunable energy level. Quite generally,
 the heat asymmetries vanish with an increasing amount of inelasticity or dephasing.
 
Our results are independent of the microscopic origin of incoherent scattering. Qualitatively,
we believe that our main conclusions will be robust and applicable to a large variety of systems.
Yet, it would be highly desirable to investigate in future works specific models taking into account, e.g.,
electron-phonon interactions.
 
 Further extensions of the model should consider cooling effects~\cite{gal09}, whose efficiency in the nonlinear regime and in the presence
 of incoherent scattering remains an open issue. Another interesting question, perhaps more fundamental, is the development of magnetic-field
 asymmetries in multiterminal setups~\cite{matthews}. It is well known that in the nonlinear regime departures of the Onsager reciprocity
 are quite general~\cite{hwa13}. The role of inelasticity and decoherence is less clear. Finally, we would like to mention the exciting possibility
 of implementing rectifying nanojunctions for energy harvesting~\cite{sot14}. A deep study of the combined effect of nonlinearities and incoherent scattering would bring the goal of waste-heat--to--electricity nanoconverters closer to reality.

\section*{Acknowledgements}
We thank Y. Apertet, J. C. Cuevas, G. Rossell\'{o} and J. Moreno for fruitful discussions. This work has been supported by a SURF@IFISC fellowship,
the MINECO under Grant No.~FIS2011-23526, the Conselleria d'Educaci\'o, Cultura i Universitats (CAIB) and FEDER.

\appendix*

\section{Charge and heat current integrals}
In our numerical analysis, it is worth to calculate the integral for the charge current through the source electrode,
\begin{equation}
\mathcal{I}_1=\frac{2e}{h}\int_\mathbb{R}dE\frac{\Gamma_1 \Gamma_2}{\pa{E-\epsilon}^2+\Gamma^2/4} [f_{1}(E)-f_{2}(E)]\,,
\label{integralCurrent}
\end{equation}
and the corresponding heat flux,
\begin{align} 
\mathcal{J}_1&=\frac{2}{h}\int_\mathbb{R}dE \pa{E-\mu_1} \frac{\Gamma_1 \Gamma_2}{\pa{E-\epsilon}^2+\Gamma^2/4}\nonumber\\
&\times [f_{1}(E)-f_{2}(E)]\,.
\label{integralHeat}
\end{align}
Here, $\epsilon$ can be $\varepsilon_0$ for the linear response or $\varepsilon_0+e\mathcal{U}(V,\theta)$
in the nonlinear regime of transport, with $\mathcal{U}$ evaluated self-consistently in terms of the applied
voltage $V$ and temperature difference $\theta$.

An analytical solution of Eqs.~\eqref{integralCurrent} and~\eqref{integralHeat} can be obtained by noticing
that the Fermi functions can be expressed in terms of the digamma function $\Psi(z)=\Gamma'(z)/\Gamma \pa{z}$:
\begin{align}
f_{j}(E)&=\frac{1}{2}\co{1-\tanh{\pa{\frac{E-\mu_{j}}{2k_BT_j}}}} \nonumber\\
&=\frac{1}{2}\co{1+\frac{i}{\pi}\co{\Psi\pa{\frac{1}{2}+i\frac{w_j(E)}{\pi}}-\Psi\pa{\frac{1}{2}-i\frac{w_j(E)}{\pi}}}}\,,
\label{fermi}
\end{align}
where $w_j(E)=(E-\mu_{j})/(2k_BT_j)$. The first (second) $\Psi$ has singularities at 
$w_j=i\pi \pa{n+\frac{1}{2}}$ [$w_j=-i\pi \pa{n+\frac{1}{2}}$] with
$n\in\mathds{N}$.
\begin{figure}
\begin{center}
\hspace*{1cm}
\includegraphics[width=0.4\textwidth]{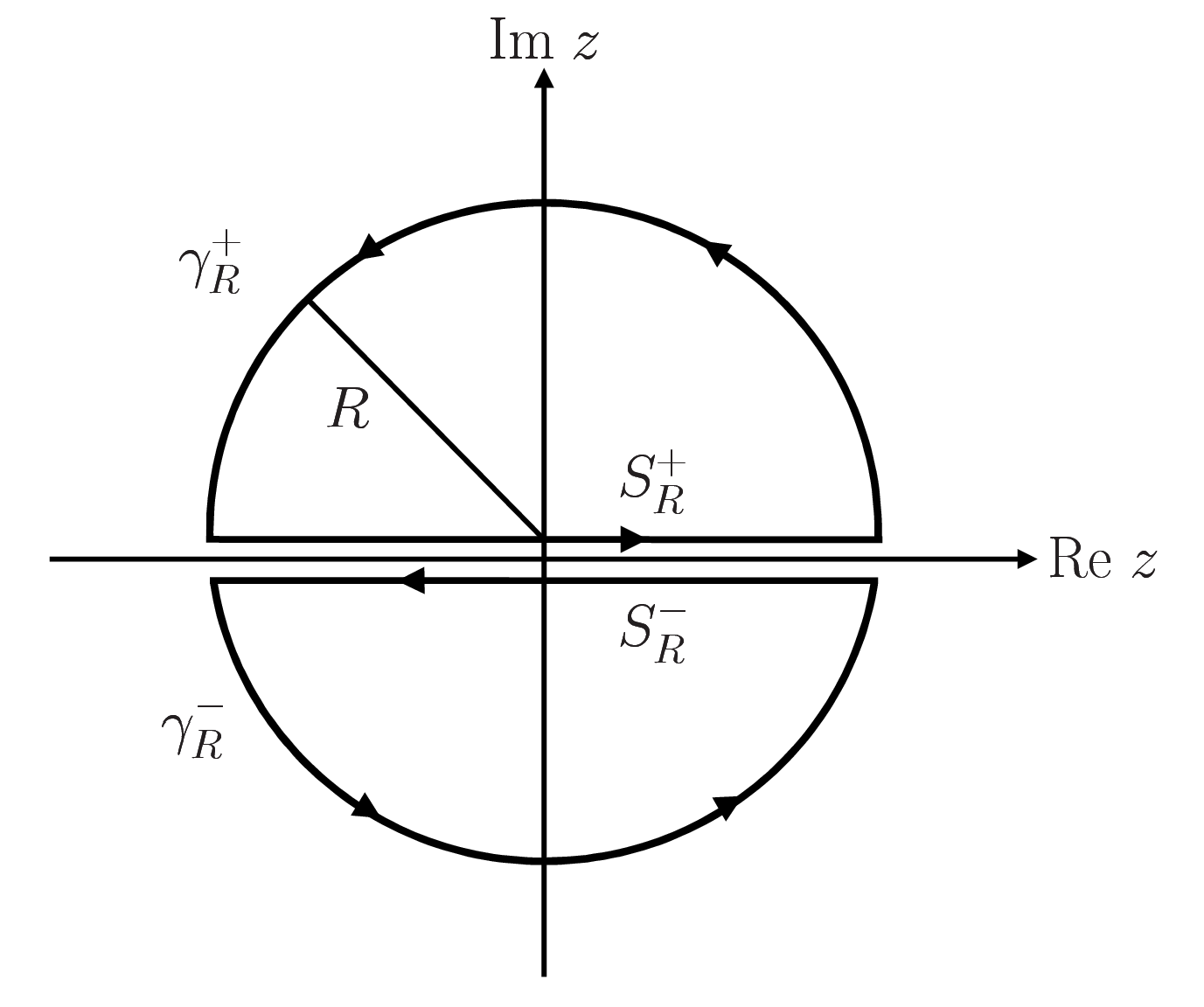}
\caption{Integration contour in the complex plane.}
\label{Fig:app}
\end{center} 
\end{figure}

Consider now the integral
\begin{equation}\label{eq_appI}
I=\int_\mathbb{R}dE \,  \tau(E) f_j(E)=I_A+I_B+I_C\,,
\end{equation}
where $\tau(E)=\co{\pa{E-\epsilon}^2+\pa{\Gamma /2}^2}^{-1}$ and
\begin{align}
I_A &=\frac{1}{2}\int_\mathbb{R}dE\,\tau(E)\,,\\
I_B &= \frac{i}{2\pi}\int_\mathbb{R}dE\,\tau(E)\Psi\pa{\frac{1}{2}+i\frac{E-\mu_j}{2\pi k_BT_j}}\,,\\
I_C &=-\frac{i}{2\pi}\int_\mathbb{R}dE\,\tau(E)\Psi\pa{\frac{1}{2}-i\frac{E-\mu_j}{2\pi k_BT_j}}\,.
\end{align}
We compute these integrals using the residue theorem. For $I_A$ and $I_C$ we choose the upper semidisk $S_R^+$ of radius $R$
while for $I_B$ it is convenient to integrate over the lower semidisk $S_R^-$ (see Fig.~\ref{Fig:app}). In the limit of infinite radius ($R\to\infty$),
the integrals along the external paths $\gamma_R^\pm$ vanish and we find
\begin{align}
I_A &=\frac{\pi}{\Gamma}\,,\\
I_B &=\frac{i}{\Gamma}\Psi\pa{\frac{1}{2}+i\frac{\epsilon-\mu_j-i\Gamma/2}{2\pi k_B T_j}}\,,\\
I_C &=-\frac{i}{\Gamma}\Psi\pa{\frac{1}{2}-i\frac{\epsilon-\mu_j+i\Gamma/2}{2\pi k_B T_j}}\,.
\end{align}
Substituting in Eq.~\eqref{eq_appI} and defining $z_j^{\pm}=\pa{\frac{1}{2}+\frac{\Gamma}{4\pi k_B T_j} \pm i\frac{\epsilon-\mu_j}{2\pi k_B T_j}}$,
Eq.~\eqref{integralCurrent} becomes
\begin{equation}
\mathcal{I}_1
=-\frac{4e}{h}\frac{\Gamma_1\Gamma_2}{\Gamma}\Im\co{\Psi\pa{z_1^{+}}
-\Psi\pa{z_2^{+}}
}\,,
\end{equation}					
where we have used the property $\Psi\pa{z^*}=\Psi^*\pa{z}$.

The expression for the heat current [Eq.~\eqref{integralHeat}] is to be treated with caution because
in this case there arises a nonzero contribution from the integration
along $\gamma_R^\pm$. Consider
\begin{align}
J &=\int_\mathbb{R} dz \pa{z-\mu_1}\tau (z)\co{f_1(z)-f_2(z)}\nonumber\\
&=\lim_{R\to\infty}\frac{i}{2\pi}({J_{\gamma_R}}-{J_S})\,,
\end{align}
with
\begin{align}
J_S &=\int_{S_R^-} dz \pa{z-\mu_1}\tau (z) \co{\Psi_1^{+}\pa{z}
-\Psi_2^{+}\pa{z}} \nonumber\\
&+\int_{S_{R}^+} dz \pa{z-\mu_1}\tau (z)\co{\Psi_1^{-}\pa{z}
-\Psi_2^{-}\pa{z}} \,,
\end{align}
\begin{align}
{J_{\gamma_R}} &= \int_{\gamma_{R}^-}dz \pa{z-\mu_1}\tau (z)\co{\Psi_1^{+}\pa{z}
-\Psi_2^{+}\pa{z}} \nonumber\\
&+\int_{\gamma_{R}^+} dz \pa{z-\mu_1}\tau (z)\co{\Psi_1^{-}\pa{z}
-\Psi_2^{-}\pa{z}}\,.
\end{align}
Here, $\Psi_j^{\pm}\pa{z}=\Psi\pa{\frac{1}{2}\pm i\frac{z-\mu_j}{2\pi k_B T_j}}$ and the paths $S_R^{\pm}$, $\gamma_R^{\pm}$ are crossed anticlockwise. $J_S$ can be obtained analogously to the charge current case:
\begin{align}
J_S &= \frac{4\pi i}{\Gamma}\pa{\epsilon-\mu_1}
\Im\co{\Psi_1^{+}\pa{z}
-\Psi_2^{+}\pa{z}} \nonumber\\
&- 2\pi i
\Re\co{\Psi_1^{+}\pa{z}
-\Psi_2^{+}\pa{z}}\,.
\label{intSemidisk}
\end{align}
To compute ${J_{\gamma_R}}=J_{\gamma_R^-}+J_{\gamma_R^+}$, we use the polar representation $z=Re^{i\theta}$,
\begin{align}\label{eq_Igamma}
J_{{\gamma}_R^{\pm}}&=
\int_{0}^{\pi} d\theta \frac{iR e^{i\theta} \pa{R e^{i\theta}\mp \mu_1}}
{\pa{R e^{i\theta}\mp \epsilon}^2+\Gamma^2/4}\nonumber\\
&\times \co{\Psi\pa{\frac{1}{2}+ i\frac{- R e^{i\theta}\pm \mu_1}{2\pi k_B T_1}}-\Psi\pa{\frac{1}{2}+ i\frac{- R e^{i\theta}\pm \mu_2}{2\pi k_BT_2}}}\,.
\end{align}
Let $g_{\pm}\pa{R,\theta}$ be the function under the integral of Eq.~\eqref{eq_Igamma}. For $\ba{z} \to \infty$ we use the asymptotic value $\Psi \pa{z} \to \ln\pa{z}$.
As a consequence, for every $\theta\in\pa{0,\pi}$ one has $
g_{\pm}\pa{R,\theta} \underset{R \to \infty}{\longrightarrow}i\ln{(T_2/T_1)}$. Thus,
\begin{equation}
\lim_{R\to\infty}J_{{\gamma}_R^{\pm}}=\pi i\ln\frac{T_2}{T_1}.
\label{intDesarroll}
\end{equation}
Collecting Eqs.~\eqref{intSemidisk} and~\eqref{intDesarroll} in Eq.~\eqref{integralHeat},
we finally obtain the analytical expression for the heat current:
\begin{align}
\mathcal{J}_1&=-\frac{4\Gamma_1 \Gamma_2}{h\Gamma}\pa{\epsilon-\mu_1}
\Im\co{\Psi\pa{z_1^{+}}
-\Psi\pa{z_2^{+}}} \nonumber \\
&+\frac{2\Gamma_1 \Gamma_2}{h}
\Re\co{\Psi\pa{z_1^{+}}-\Psi\pa{z_2^{+}}}-\frac{2\Gamma_1 \Gamma_2}{h}\ln\frac{T_2}{T_1}.
\end{align}

\bibliography{biblio} 

\begin{thebibliography}{63}
\expandafter\ifx\csname natexlab\endcsname\relax\def\natexlab#1{#1}\fi
\expandafter\ifx\csname bibnamefont\endcsname\relax
  \def\bibnamefont#1{#1}\fi
\expandafter\ifx\csname bibfnamefont\endcsname\relax
  \def\bibfnamefont#1{#1}\fi
\expandafter\ifx\csname citenamefont\endcsname\relax
  \def\citenamefont#1{#1}\fi
\expandafter\ifx\csname url\endcsname\relax
  \def\url#1{\texttt{#1}}\fi
\expandafter\ifx\csname urlprefix\endcsname\relax\def\urlprefix{URL }\fi
\providecommand{\bibinfo}[2]{#2}
\providecommand{\eprint}[2][]{\url{#2}}

\bibitem[{\citenamefont{S{\'a}nchez and Linke}(2014)}]{san14}
\bibinfo{author}{\bibfnamefont{D.}~\bibnamefont{S{\'a}nchez}} \bibnamefont{and}
  \bibinfo{author}{\bibfnamefont{H.}~\bibnamefont{Linke}},
  \emph{\bibinfo{title}{Focus on thermoelectric effects in nanostructures}},
  \bibinfo{journal}{New. J. Phys.} \textbf{\bibinfo{volume}{16}},
  \bibinfo{pages}{110201} (\bibinfo{year}{2014}), \bibinfo{note}{and references
  cited therein}.

\bibitem[{\citenamefont{Kim et~al.}(2012)\citenamefont{Kim, Jeong, Lee, and
  Reddy}}]{kim12}
\bibinfo{author}{\bibfnamefont{K.}~\bibnamefont{Kim}},
  \bibinfo{author}{\bibfnamefont{W.}~\bibnamefont{Jeong}},
  \bibinfo{author}{\bibfnamefont{W.}~\bibnamefont{Lee}}, \bibnamefont{and}
  \bibinfo{author}{\bibfnamefont{P.}~\bibnamefont{Reddy}},
  \emph{\bibinfo{title}{Ultra-high vacuum scanning thermal microscopy for
  nanometer resolution quantitative thermometry}}, \bibinfo{journal}{ACS Nano}
  \textbf{\bibinfo{volume}{6}}, \bibinfo{pages}{4248} (\bibinfo{year}{2012}).

\bibitem[{\citenamefont{Shakouri}(2006)}]{Sha06}
\bibinfo{author}{\bibfnamefont{A.}~\bibnamefont{Shakouri}},
  \emph{\bibinfo{title}{Nanoscale thermal transport and microrefrigerators on a
  chip}}, \bibinfo{journal}{IEEE} \textbf{\bibinfo{volume}{94}},
  \bibinfo{pages}{1613} (\bibinfo{year}{2006}).

\bibitem[{\citenamefont{Giazotto et~al.}(2006)\citenamefont{Giazotto,
  Heikkil{\"a}, Luukanen, Savin, and Pekola}}]{Gia06}
\bibinfo{author}{\bibfnamefont{F.}~\bibnamefont{Giazotto}},
  \bibinfo{author}{\bibfnamefont{T.~T.} \bibnamefont{Heikkil{\"a}}},
  \bibinfo{author}{\bibfnamefont{A.}~\bibnamefont{Luukanen}},
  \bibinfo{author}{\bibfnamefont{A.~M.} \bibnamefont{Savin}}, \bibnamefont{and}
  \bibinfo{author}{\bibfnamefont{J.~P.} \bibnamefont{Pekola}},
  \emph{\bibinfo{title}{Opportunities for mesoscopics in thermometry and
  refrigeration: Physics and applications}}, \bibinfo{journal}{Rev. Mod. Phys.}
  \textbf{\bibinfo{volume}{78}}, \bibinfo{pages}{217} (\bibinfo{year}{2006}).

\bibitem[{\citenamefont{Terraneo et~al.}(2002)\citenamefont{Terraneo, Peyrard,
  and Casati}}]{Terr02}
\bibinfo{author}{\bibfnamefont{M.}~\bibnamefont{Terraneo}},
  \bibinfo{author}{\bibfnamefont{M.}~\bibnamefont{Peyrard}}, \bibnamefont{and}
  \bibinfo{author}{\bibfnamefont{G.}~\bibnamefont{Casati}},
  \emph{\bibinfo{title}{Controlling the energy flow in nonlinear lattices: a
  model for a thermal rectifier}}, \bibinfo{journal}{Phys. Rev. Lett.}
  \textbf{\bibinfo{volume}{88}}, \bibinfo{pages}{094302}
  (\bibinfo{year}{2002}).

\bibitem[{\citenamefont{Heremans et~al.}(2013)\citenamefont{Heremans,
  Dresselhaus, Bell, and Morelli}}]{her13}
\bibinfo{author}{\bibfnamefont{J.~P.} \bibnamefont{Heremans}},
  \bibinfo{author}{\bibfnamefont{M.~S.} \bibnamefont{Dresselhaus}},
  \bibinfo{author}{\bibfnamefont{L.~E.} \bibnamefont{Bell}}, \bibnamefont{and}
  \bibinfo{author}{\bibfnamefont{D.~T.} \bibnamefont{Morelli}},
  \emph{\bibinfo{title}{When thermoelectrics reached the nanoscale}},
  \bibinfo{journal}{Nature Nanotech.} \textbf{\bibinfo{volume}{8}},
  \bibinfo{pages}{471} (\bibinfo{year}{2013}).

\bibitem[{\citenamefont{Snyder and Toberer}(2008)}]{Jeff07}
\bibinfo{author}{\bibfnamefont{G.~J.} \bibnamefont{Snyder}} \bibnamefont{and}
  \bibinfo{author}{\bibfnamefont{E.~S.} \bibnamefont{Toberer}},
  \emph{\bibinfo{title}{Complex thermoelectric materials}},
  \bibinfo{journal}{Nature Mater.} \textbf{\bibinfo{volume}{7}},
  \bibinfo{pages}{105} (\bibinfo{year}{2008}).

\bibitem[{\citenamefont{Li et~al.}(2010)\citenamefont{Li, Liu, Zhao, and
  Zhou}}]{npg}
\bibinfo{author}{\bibfnamefont{J.-F.} \bibnamefont{Li}},
  \bibinfo{author}{\bibfnamefont{W.-S.} \bibnamefont{Liu}},
  \bibinfo{author}{\bibfnamefont{L.-D.} \bibnamefont{Zhao}}, \bibnamefont{and}
  \bibinfo{author}{\bibfnamefont{M.}~\bibnamefont{Zhou}},
  \emph{\bibinfo{title}{High-performance nanostructured thermoelectric
  materials}}, \bibinfo{journal}{NPG Asia Materials}
  \textbf{\bibinfo{volume}{2}}, \bibinfo{pages}{152} (\bibinfo{year}{2010}).

\bibitem[{\citenamefont{Jezouin et~al.}(2013)\citenamefont{Jezouin, Parmentier,
  Anthore, Gennser, Cavanna, Jin, and Pierre}}]{jez13}
\bibinfo{author}{\bibfnamefont{S.}~\bibnamefont{Jezouin}},
  \bibinfo{author}{\bibfnamefont{F.~D.} \bibnamefont{Parmentier}},
  \bibinfo{author}{\bibfnamefont{A.}~\bibnamefont{Anthore}},
  \bibinfo{author}{\bibfnamefont{U.}~\bibnamefont{Gennser}},
  \bibinfo{author}{\bibfnamefont{A.}~\bibnamefont{Cavanna}},
  \bibinfo{author}{\bibfnamefont{Y.}~\bibnamefont{Jin}}, \bibnamefont{and}
  \bibinfo{author}{\bibfnamefont{F.}~\bibnamefont{Pierre}},
  \emph{\bibinfo{title}{Quantum limit of heat flow across a single electronic
  channel}}, \bibinfo{journal}{Science} \textbf{\bibinfo{volume}{342}},
  \bibinfo{pages}{601} (\bibinfo{year}{2013}).

\bibitem[{\citenamefont{Molenkamp et~al.}(1992)\citenamefont{Molenkamp,
  Gravier, van Houten, Buijk, Mabesoone, and Foxon}}]{mol92}
\bibinfo{author}{\bibfnamefont{L.~W.} \bibnamefont{Molenkamp}},
  \bibinfo{author}{\bibfnamefont{T.}~\bibnamefont{Gravier}},
  \bibinfo{author}{\bibfnamefont{H.}~\bibnamefont{van Houten}},
  \bibinfo{author}{\bibfnamefont{O.~J.~A.} \bibnamefont{Buijk}},
  \bibinfo{author}{\bibfnamefont{M.~A.~A.} \bibnamefont{Mabesoone}},
  \bibnamefont{and} \bibinfo{author}{\bibfnamefont{C.~T.} \bibnamefont{Foxon}},
  \emph{\bibinfo{title}{Peltier coefficient and thermal conductance of a
  quantum point contact}}, \bibinfo{journal}{Phys. Rev. Lett.}
  \textbf{\bibinfo{volume}{68}}, \bibinfo{pages}{3765} (\bibinfo{year}{1992}).

\bibitem[{\citenamefont{Kane et~al.}(1997)\citenamefont{Kane, Balents, and
  Fisher}}]{kan97}
\bibinfo{author}{\bibfnamefont{C.}~\bibnamefont{Kane}},
  \bibinfo{author}{\bibfnamefont{L.}~\bibnamefont{Balents}}, \bibnamefont{and}
  \bibinfo{author}{\bibfnamefont{M.~P.~A.} \bibnamefont{Fisher}},
  \emph{\bibinfo{title}{Coulomb interactions and mesoscopic effects in carbon
  nanotubes}}, \bibinfo{journal}{Phys. Rev. Lett.}
  \textbf{\bibinfo{volume}{79}}, \bibinfo{pages}{5086} (\bibinfo{year}{1997}).

\bibitem[{\citenamefont{Battista et~al.}(2013)\citenamefont{Battista,
  Moskalets, Albert, and Samuelsson}}]{bat13}
\bibinfo{author}{\bibfnamefont{F.}~\bibnamefont{Battista}},
  \bibinfo{author}{\bibfnamefont{M.}~\bibnamefont{Moskalets}},
  \bibinfo{author}{\bibfnamefont{M.}~\bibnamefont{Albert}}, \bibnamefont{and}
  \bibinfo{author}{\bibfnamefont{P.}~\bibnamefont{Samuelsson}},
  \emph{\bibinfo{title}{Quantum heat fluctuations of single-particle sources}},
  \bibinfo{journal}{Phys. Rev. Lett.} \textbf{\bibinfo{volume}{110}},
  \bibinfo{pages}{126602} (\bibinfo{year}{2013}).

\bibitem[{\citenamefont{Spilla et~al.}(2014)\citenamefont{Spilla, Hassler, and
  Splettstoesser}}]{spi14}
\bibinfo{author}{\bibfnamefont{S.}~\bibnamefont{Spilla}},
  \bibinfo{author}{\bibfnamefont{F.}~\bibnamefont{Hassler}}, \bibnamefont{and}
  \bibinfo{author}{\bibfnamefont{J.}~\bibnamefont{Splettstoesser}},
  \emph{\bibinfo{title}{Measurement and dephasing of a flux qubit due to heat
  currents}}, \bibinfo{journal}{New. J. Phys.} \textbf{\bibinfo{volume}{16}},
  \bibinfo{pages}{045020} (\bibinfo{year}{2014}).

\bibitem[{\citenamefont{Lee et~al.}(2013)\citenamefont{Lee, Kim, Jeong, Zotti,
  Pauly, Cuevas, and Reddy}}]{Lee13}
\bibinfo{author}{\bibfnamefont{W.}~\bibnamefont{Lee}},
  \bibinfo{author}{\bibfnamefont{K.}~\bibnamefont{Kim}},
  \bibinfo{author}{\bibfnamefont{W.}~\bibnamefont{Jeong}},
  \bibinfo{author}{\bibfnamefont{L.~A.} \bibnamefont{Zotti}},
  \bibinfo{author}{\bibfnamefont{F.}~\bibnamefont{Pauly}},
  \bibinfo{author}{\bibfnamefont{J.~C.} \bibnamefont{Cuevas}},
  \bibnamefont{and} \bibinfo{author}{\bibfnamefont{P.}~\bibnamefont{Reddy}},
  \emph{\bibinfo{title}{Heat dissipation in atomic-scale junctions}},
  \bibinfo{journal}{Nature (London)} \textbf{\bibinfo{volume}{498}},
  \bibinfo{pages}{209} (\bibinfo{year}{2013}).

\bibitem[{\citenamefont{Zotti et~al.}(2014)\citenamefont{Zotti, B{\"u}rkle,
  Pauly, Lee, Kim, Jeong, Asai, Reddy, and Cuevas}}]{zot14}
\bibinfo{author}{\bibfnamefont{L.~A.} \bibnamefont{Zotti}},
  \bibinfo{author}{\bibfnamefont{M.}~\bibnamefont{B{\"u}rkle}},
  \bibinfo{author}{\bibfnamefont{F.}~\bibnamefont{Pauly}},
  \bibinfo{author}{\bibfnamefont{W.}~\bibnamefont{Lee}},
  \bibinfo{author}{\bibfnamefont{K.}~\bibnamefont{Kim}},
  \bibinfo{author}{\bibfnamefont{W.}~\bibnamefont{Jeong}},
  \bibinfo{author}{\bibfnamefont{Y.}~\bibnamefont{Asai}},
  \bibinfo{author}{\bibfnamefont{P.}~\bibnamefont{Reddy}}, \bibnamefont{and}
  \bibinfo{author}{\bibfnamefont{J.~C.} \bibnamefont{Cuevas}},
  \emph{\bibinfo{title}{Heat dissipation and its relation to thermopower in
  single-molecule junctions}}, \bibinfo{journal}{New. J. Phys.}
  \textbf{\bibinfo{volume}{16}}, \bibinfo{pages}{015004}
  (\bibinfo{year}{2014}).

\bibitem[{\citenamefont{Paulsson and Datta}(2003)}]{pau03}
\bibinfo{author}{\bibfnamefont{M.}~\bibnamefont{Paulsson}} \bibnamefont{and}
  \bibinfo{author}{\bibfnamefont{S.}~\bibnamefont{Datta}},
  \emph{\bibinfo{title}{Thermoelectric effect in molecular electronics}},
  \bibinfo{journal}{Phys. Rev. B} \textbf{\bibinfo{volume}{67}},
  \bibinfo{pages}{241403} (\bibinfo{year}{2003}).

\bibitem[{\citenamefont{Frederiksen et~al.}(2004)\citenamefont{Frederiksen,
  Brandbyge, Lorente, and Jauho}}]{fre04}
\bibinfo{author}{\bibfnamefont{T.}~\bibnamefont{Frederiksen}},
  \bibinfo{author}{\bibfnamefont{M.}~\bibnamefont{Brandbyge}},
  \bibinfo{author}{\bibfnamefont{N.}~\bibnamefont{Lorente}}, \bibnamefont{and}
  \bibinfo{author}{\bibfnamefont{A.-P.} \bibnamefont{Jauho}},
  \emph{\bibinfo{title}{Inelastic scattering and local heating in atomic gold
  wires}}, \bibinfo{journal}{Phys. Rev. Lett.} \textbf{\bibinfo{volume}{93}},
  \bibinfo{pages}{256601} (\bibinfo{year}{2004}).

\bibitem[{\citenamefont{Koch et~al.}(2004)\citenamefont{Koch, von Oppen, Oreg,
  and Sela}}]{koc04}
\bibinfo{author}{\bibfnamefont{J.}~\bibnamefont{Koch}},
  \bibinfo{author}{\bibfnamefont{F.}~\bibnamefont{von Oppen}},
  \bibinfo{author}{\bibfnamefont{Y.}~\bibnamefont{Oreg}}, \bibnamefont{and}
  \bibinfo{author}{\bibfnamefont{E.}~\bibnamefont{Sela}},
  \emph{\bibinfo{title}{Thermopower of single-molecule devices}},
  \bibinfo{journal}{Phys. Rev. B} \textbf{\bibinfo{volume}{70}},
  \bibinfo{pages}{195107} (\bibinfo{year}{2004}).

\bibitem[{\citenamefont{Galperin et~al.}(2007)\citenamefont{Galperin, Nitzan,
  and Ratner}}]{gal07}
\bibinfo{author}{\bibfnamefont{M.}~\bibnamefont{Galperin}},
  \bibinfo{author}{\bibfnamefont{A.}~\bibnamefont{Nitzan}}, \bibnamefont{and}
  \bibinfo{author}{\bibfnamefont{M.~A.} \bibnamefont{Ratner}},
  \emph{\bibinfo{title}{Heat conduction in molecular transport junctions}},
  \bibinfo{journal}{Phys. Rev. B} \textbf{\bibinfo{volume}{75}},
  \bibinfo{pages}{155312} (\bibinfo{year}{2007}).

\bibitem[{\citenamefont{Finch et~al.}(2009)\citenamefont{Finch,
  Garc{\'i}a-Su{\'a}rez, and Lambert}}]{fin09}
\bibinfo{author}{\bibfnamefont{C.~M.} \bibnamefont{Finch}},
  \bibinfo{author}{\bibfnamefont{V.~M.} \bibnamefont{Garc{\'i}a-Su{\'a}rez}},
  \bibnamefont{and} \bibinfo{author}{\bibfnamefont{C.~J.}
  \bibnamefont{Lambert}}, \emph{\bibinfo{title}{Giant thermopower and figure of
  merit in single-molecule devices}}, \bibinfo{journal}{Phys. Rev. B}
  \textbf{\bibinfo{volume}{79}}, \bibinfo{pages}{033405}
  (\bibinfo{year}{2009}).

\bibitem[{\citenamefont{Entin-Wohlman and Aharony}(2012)}]{Entin12}
\bibinfo{author}{\bibfnamefont{O.}~\bibnamefont{Entin-Wohlman}}
  \bibnamefont{and} \bibinfo{author}{\bibfnamefont{A.}~\bibnamefont{Aharony}},
  \emph{\bibinfo{title}{Three-terminal thermoelectric transport under broken
  time-reversal symmetry}}, \bibinfo{journal}{Phys. Rev. B}
  \textbf{\bibinfo{volume}{85}}, \bibinfo{pages}{085401}
  (\bibinfo{year}{2012}).

\bibitem[{\citenamefont{Zimbovskaya}(2014)}]{zim14}
\bibinfo{author}{\bibfnamefont{N.~A.} \bibnamefont{Zimbovskaya}},
  \emph{\bibinfo{title}{The effect of dephasing on the thermoelectric
  efficiency of molecular junctions}}, \bibinfo{journal}{J. Phys.: Condens.
  Matter} \textbf{\bibinfo{volume}{26}}, \bibinfo{pages}{275303}
  (\bibinfo{year}{2014}).

\bibitem[{\citenamefont{B{\"u}ttiker}(1988)}]{Butt88}
\bibinfo{author}{\bibfnamefont{M.}~\bibnamefont{B{\"u}ttiker}},
  \emph{\bibinfo{title}{Coherent and sequential tunneling in series barriers}},
  \bibinfo{journal}{IBM J. Res. Dev.} \textbf{\bibinfo{volume}{32}},
  \bibinfo{pages}{63} (\bibinfo{year}{1988}).

\bibitem[{\citenamefont{D'Amato and Pastawski}(1990)}]{ama90}
\bibinfo{author}{\bibfnamefont{J.~L.} \bibnamefont{D'Amato}} \bibnamefont{and}
  \bibinfo{author}{\bibfnamefont{H.~M.} \bibnamefont{Pastawski}},
  \emph{\bibinfo{title}{Conductance of a disordered linear chain including
  inelastic scattering events}}, \bibinfo{journal}{Phys. Rev. B}
  \textbf{\bibinfo{volume}{41}}, \bibinfo{pages}{7411} (\bibinfo{year}{1990}).

\bibitem[{\citenamefont{de~Jong and Beenakker}(1996)}]{jon96}
\bibinfo{author}{\bibfnamefont{M.~J.~M.} \bibnamefont{de~Jong}}
  \bibnamefont{and} \bibinfo{author}{\bibfnamefont{C.~W.~J.}
  \bibnamefont{Beenakker}}, \emph{\bibinfo{title}{Semiclassical theory of shot
  noise in mesoscopic conductors}}, \bibinfo{journal}{Physica A}
  \textbf{\bibinfo{volume}{230}}, \bibinfo{pages}{219} (\bibinfo{year}{1996}).

\bibitem[{\citenamefont{van Langen and B{\"u}ttiker}(1997)}]{lan97}
\bibinfo{author}{\bibfnamefont{S.~A.} \bibnamefont{van Langen}}
  \bibnamefont{and}
  \bibinfo{author}{\bibfnamefont{M.}~\bibnamefont{B{\"u}ttiker}},
  \emph{\bibinfo{title}{Quantum-statistical current correlations in multilead
  chaotic cavities}}, \bibinfo{journal}{Phys. Rev. B}
  \textbf{\bibinfo{volume}{56}}, \bibinfo{pages}{R1680} (\bibinfo{year}{1997}).

\bibitem[{\citenamefont{Saito et~al.}(2011)\citenamefont{Saito, Benenti,
  Casati, and Prosen}}]{Saito11}
\bibinfo{author}{\bibfnamefont{K.}~\bibnamefont{Saito}},
  \bibinfo{author}{\bibfnamefont{G.}~\bibnamefont{Benenti}},
  \bibinfo{author}{\bibfnamefont{G.}~\bibnamefont{Casati}}, \bibnamefont{and}
  \bibinfo{author}{\bibfnamefont{T.}~\bibnamefont{Prosen}},
  \emph{\bibinfo{title}{Thermopower with broken time-reversal symmetry}},
  \bibinfo{journal}{Phys. Rev. B} \textbf{\bibinfo{volume}{84}},
  \bibinfo{pages}{201306} (\bibinfo{year}{2011}).

\bibitem[{\citenamefont{S{\'a}nchez and Serra}(2011)}]{dav11}
\bibinfo{author}{\bibfnamefont{D.}~\bibnamefont{S{\'a}nchez}} \bibnamefont{and}
  \bibinfo{author}{\bibfnamefont{L.}~\bibnamefont{Serra}},
  \emph{\bibinfo{title}{Thermoelectric transport of mesoscopic conductors
  coupled to voltage and thermal probes}}, \bibinfo{journal}{Phys. Rev. B}
  \textbf{\bibinfo{volume}{84}}, \bibinfo{pages}{201307}
  (\bibinfo{year}{2011}).

\bibitem[{\citenamefont{Caso et~al.}(2012)\citenamefont{Caso, Arrachea, and
  Lozano}}]{Caso12}
\bibinfo{author}{\bibfnamefont{A.}~\bibnamefont{Caso}},
  \bibinfo{author}{\bibfnamefont{L.}~\bibnamefont{Arrachea}}, \bibnamefont{and}
  \bibinfo{author}{\bibfnamefont{G.~S.} \bibnamefont{Lozano}},
  \emph{\bibinfo{title}{Defining the effective temperature of a quantum driven
  system from current-current correlation functions}}, \bibinfo{journal}{Eur.
  Phys. J. B} \textbf{\bibinfo{volume}{85}}, \bibinfo{pages}{1}
  (\bibinfo{year}{2012}).

\bibitem[{\citenamefont{Bedkihal et~al.}(2013)\citenamefont{Bedkihal,
  Bandyopadhyay, and Segal}}]{bed13}
\bibinfo{author}{\bibfnamefont{S.}~\bibnamefont{Bedkihal}},
  \bibinfo{author}{\bibfnamefont{M.}~\bibnamefont{Bandyopadhyay}},
  \bibnamefont{and} \bibinfo{author}{\bibfnamefont{D.}~\bibnamefont{Segal}},
  \emph{\bibinfo{title}{The probe technique far from equilibrium: Magnetic
  field symmetries of nonlinear transport}}, \bibinfo{journal}{Eur. Phys. J. B}
  \textbf{\bibinfo{volume}{86}}, \bibinfo{pages}{1} (\bibinfo{year}{2013}).

\bibitem[{\citenamefont{Bergfield et~al.}(2013)\citenamefont{Bergfield, Story,
  Stafford, and Stafford}}]{ber13}
\bibinfo{author}{\bibfnamefont{J.~P.} \bibnamefont{Bergfield}},
  \bibinfo{author}{\bibfnamefont{S.~M.} \bibnamefont{Story}},
  \bibinfo{author}{\bibfnamefont{R.~C.} \bibnamefont{Stafford}},
  \bibnamefont{and} \bibinfo{author}{\bibfnamefont{C.~A.}
  \bibnamefont{Stafford}}, \emph{\bibinfo{title}{Probing {Maxwell's} demon with
  a nanoscale thermometer}}, \bibinfo{journal}{ACS Nano}
  \textbf{\bibinfo{volume}{7}}, \bibinfo{pages}{4429} (\bibinfo{year}{2013}).

\bibitem[{\citenamefont{Apertet et~al.}(2013)\citenamefont{Apertet, Ouerdane,
  Goupil, and Lecoeur}}]{ape13}
\bibinfo{author}{\bibfnamefont{Y.}~\bibnamefont{Apertet}},
  \bibinfo{author}{\bibfnamefont{H.}~\bibnamefont{Ouerdane}},
  \bibinfo{author}{\bibfnamefont{C.}~\bibnamefont{Goupil}}, \bibnamefont{and}
  \bibinfo{author}{\bibfnamefont{P.}~\bibnamefont{Lecoeur}},
  \emph{\bibinfo{title}{From local force-flux relationships to internal
  dissipations and their impact on heat engine performance: The illustrative
  case of a thermoelectric generator}}, \bibinfo{journal}{Phys. Rev. E}
  \textbf{\bibinfo{volume}{88}}, \bibinfo{pages}{022137}
  (\bibinfo{year}{2013}).

\bibitem[{\citenamefont{Brandner and Seifert}(2013)}]{bra13}
\bibinfo{author}{\bibfnamefont{K.}~\bibnamefont{Brandner}} \bibnamefont{and}
  \bibinfo{author}{\bibfnamefont{U.}~\bibnamefont{Seifert}},
  \emph{\bibinfo{title}{Multi-terminal thermoelectric transport in a magnetic
  field: bounds on {Onsager} coefficients and efficiency}},
  \bibinfo{journal}{New. J. Phys.} \textbf{\bibinfo{volume}{15}},
  \bibinfo{pages}{105003} (\bibinfo{year}{2013}).

\bibitem[{\citenamefont{Meair et~al.}(2014)\citenamefont{Meair, Bergfield,
  Stafford, and Jacquod}}]{Meair14}
\bibinfo{author}{\bibfnamefont{J.}~\bibnamefont{Meair}},
  \bibinfo{author}{\bibfnamefont{J.~P.} \bibnamefont{Bergfield}},
  \bibinfo{author}{\bibfnamefont{C.~A.} \bibnamefont{Stafford}},
  \bibnamefont{and} \bibinfo{author}{\bibfnamefont{P.}~\bibnamefont{Jacquod}},
  \emph{\bibinfo{title}{Local temperature of out-of-equilibrium quantum
  electron systems}}, \bibinfo{journal}{Phys. Rev. B}
  \textbf{\bibinfo{volume}{90}}, \bibinfo{pages}{035407}
  (\bibinfo{year}{2014}).

\bibitem[{\citenamefont{Butcher}(1990)}]{butcher}
\bibinfo{author}{\bibfnamefont{P.~N.} \bibnamefont{Butcher}},
  \emph{\bibinfo{title}{Thermal and electrical transport formalism for
  electronic microstructures with many terminals}}, \bibinfo{journal}{J. Phys.:
  Condens. Matter} \textbf{\bibinfo{volume}{2}}, \bibinfo{pages}{4869}
  (\bibinfo{year}{1990}).

\bibitem[{\citenamefont{Christen and B{\"u}ttiker}(1996)}]{chr96}
\bibinfo{author}{\bibfnamefont{T.}~\bibnamefont{Christen}} \bibnamefont{and}
  \bibinfo{author}{\bibfnamefont{M.}~\bibnamefont{B{\"u}ttiker}},
  \emph{\bibinfo{title}{Gauge-invariant nonlinear electric transport in
  mesoscopic conductors}}, \bibinfo{journal}{EPL}
  \textbf{\bibinfo{volume}{35}}, \bibinfo{pages}{523} (\bibinfo{year}{1996}).

\bibitem[{\citenamefont{S{\'a}nchez and L{\'o}pez}(2013)}]{san13}
\bibinfo{author}{\bibfnamefont{D.}~\bibnamefont{S{\'a}nchez}} \bibnamefont{and}
  \bibinfo{author}{\bibfnamefont{R.}~\bibnamefont{L{\'o}pez}},
  \emph{\bibinfo{title}{Scattering theory of nonlinear thermoelectric
  transport}}, \bibinfo{journal}{Phys. Rev. Lett.}
  \textbf{\bibinfo{volume}{110}}, \bibinfo{pages}{026804}
  (\bibinfo{year}{2013}).

\bibitem[{\citenamefont{Meair and Jacquod}(2013)}]{jac13}
\bibinfo{author}{\bibfnamefont{J.}~\bibnamefont{Meair}} \bibnamefont{and}
  \bibinfo{author}{\bibfnamefont{P.}~\bibnamefont{Jacquod}},
  \emph{\bibinfo{title}{Scattering theory of nonlinear thermoelectricity in
  quantum coherent conductors}}, \bibinfo{journal}{J. Phys.: Condens. Matter}
  \textbf{\bibinfo{volume}{25}}, \bibinfo{pages}{082201}
  (\bibinfo{year}{2013}).

\bibitem[{\citenamefont{Cutler and Mott}(1969)}]{mott}
\bibinfo{author}{\bibfnamefont{M.}~\bibnamefont{Cutler}} \bibnamefont{and}
  \bibinfo{author}{\bibfnamefont{N.~F.} \bibnamefont{Mott}},
  \emph{\bibinfo{title}{Observation of {Anderson} localization in an electron
  gas}}, \bibinfo{journal}{Phys. Rev.} \textbf{\bibinfo{volume}{181}},
  \bibinfo{pages}{1336} (\bibinfo{year}{1969}).

\bibitem[{\citenamefont{Jonson and Mahan}(1980)}]{jon80}
\bibinfo{author}{\bibfnamefont{M.}~\bibnamefont{Jonson}} \bibnamefont{and}
  \bibinfo{author}{\bibfnamefont{G.~D.} \bibnamefont{Mahan}},
  \emph{\bibinfo{title}{Mott's formula for the thermopower and the
  {Wiedemann-Franz} law}}, \bibinfo{journal}{Phys. Rev. B}
  \textbf{\bibinfo{volume}{21}}, \bibinfo{pages}{4223} (\bibinfo{year}{1980}).

\bibitem[{\citenamefont{Meir and Wingreen}(1992)}]{mei91}
\bibinfo{author}{\bibfnamefont{Y.}~\bibnamefont{Meir}} \bibnamefont{and}
  \bibinfo{author}{\bibfnamefont{N.~S.} \bibnamefont{Wingreen}},
  \emph{\bibinfo{title}{Landauer formula for the current through an interacting
  electron region}}, \bibinfo{journal}{Phys. Rev. Lett.}
  \textbf{\bibinfo{volume}{68}}, \bibinfo{pages}{2512} (\bibinfo{year}{1992}).

\bibitem[{\citenamefont{Staring et~al.}(1993)\citenamefont{Staring, Molenkamp,
  Alphenaar, van Houten, Buyk, Mabesoone, Beenakker, and Foxon}}]{Star93}
\bibinfo{author}{\bibfnamefont{A.~A.~M.} \bibnamefont{Staring}},
  \bibinfo{author}{\bibfnamefont{L.~W.} \bibnamefont{Molenkamp}},
  \bibinfo{author}{\bibfnamefont{B.~W.} \bibnamefont{Alphenaar}},
  \bibinfo{author}{\bibfnamefont{H.}~\bibnamefont{van Houten}},
  \bibinfo{author}{\bibfnamefont{O.~J.~A.} \bibnamefont{Buyk}},
  \bibinfo{author}{\bibfnamefont{M.~A.~A.} \bibnamefont{Mabesoone}},
  \bibinfo{author}{\bibfnamefont{C.~W.~J.} \bibnamefont{Beenakker}},
  \bibnamefont{and} \bibinfo{author}{\bibfnamefont{C.~T.} \bibnamefont{Foxon}},
  \emph{\bibinfo{title}{Coulomb-blockade oscillations in the thermopower of a
  quantum dot}}, \bibinfo{journal}{EPL} \textbf{\bibinfo{volume}{22}},
  \bibinfo{pages}{57} (\bibinfo{year}{1993}).

\bibitem[{\citenamefont{Svensson et~al.}(2013)\citenamefont{Svensson, Hoffmann,
  Nakpathomkun, Wu, Xu, Nilsson, S{\'a}nchez, Kashcheyevs, and Linke}}]{Sve13}
\bibinfo{author}{\bibfnamefont{S.~F.} \bibnamefont{Svensson}},
  \bibinfo{author}{\bibfnamefont{E.~A.} \bibnamefont{Hoffmann}},
  \bibinfo{author}{\bibfnamefont{N.}~\bibnamefont{Nakpathomkun}},
  \bibinfo{author}{\bibfnamefont{P.~M.} \bibnamefont{Wu}},
  \bibinfo{author}{\bibfnamefont{H.~Q.} \bibnamefont{Xu}},
  \bibinfo{author}{\bibfnamefont{H.~A.} \bibnamefont{Nilsson}},
  \bibinfo{author}{\bibfnamefont{D.}~\bibnamefont{S{\'a}nchez}},
  \bibinfo{author}{\bibfnamefont{V.}~\bibnamefont{Kashcheyevs}},
  \bibnamefont{and} \bibinfo{author}{\bibfnamefont{H.}~\bibnamefont{Linke}},
  \emph{\bibinfo{title}{Nonlinear thermovoltage and thermocurrent in quantum
  dots}}, \bibinfo{journal}{New. J. Phys.} \textbf{\bibinfo{volume}{15}},
  \bibinfo{pages}{105011} (\bibinfo{year}{2013}).

\bibitem[{\citenamefont{Sierra and S{\'a}nchez}(2014)}]{sie14}
\bibinfo{author}{\bibfnamefont{M.~A.} \bibnamefont{Sierra}} \bibnamefont{and}
  \bibinfo{author}{\bibfnamefont{D.}~\bibnamefont{S{\'a}nchez}},
  \emph{\bibinfo{title}{Strongly nonlinear thermovoltage and heat dissipation
  in interacting quantum dots}}, \bibinfo{journal}{Phys. Rev. B}
  \textbf{\bibinfo{volume}{90}}, \bibinfo{pages}{115313}
  (\bibinfo{year}{2014}).

\bibitem[{\citenamefont{Kulik}(1994)}]{kul94}
\bibinfo{author}{\bibfnamefont{I.~O.} \bibnamefont{Kulik}},
  \emph{\bibinfo{title}{Non-linear thermoelectricity and cooling effects in
  metallic constrictions}}, \bibinfo{journal}{J. Phys.: Condens. Matter}
  \textbf{\bibinfo{volume}{6}}, \bibinfo{pages}{9737} (\bibinfo{year}{1994}).

\bibitem[{\citenamefont{Bogachek et~al.}(1999)\citenamefont{Bogachek,
  Scherbakov, and Landman}}]{bog99}
\bibinfo{author}{\bibfnamefont{E.~N.} \bibnamefont{Bogachek}},
  \bibinfo{author}{\bibfnamefont{A.~G.} \bibnamefont{Scherbakov}},
  \bibnamefont{and} \bibinfo{author}{\bibfnamefont{U.}~\bibnamefont{Landman}},
  \emph{\bibinfo{title}{Nonlinear {Peltier} effect and thermoconductance in
  nanowires}}, \bibinfo{journal}{Phys. Rev. B} \textbf{\bibinfo{volume}{60}},
  \bibinfo{pages}{11678} (\bibinfo{year}{1999}).

\bibitem[{\citenamefont{Li et~al.}(2004)\citenamefont{Li, Wang, and
  Casati}}]{li04}
\bibinfo{author}{\bibfnamefont{B.}~\bibnamefont{Li}},
  \bibinfo{author}{\bibfnamefont{L.}~\bibnamefont{Wang}}, \bibnamefont{and}
  \bibinfo{author}{\bibfnamefont{G.}~\bibnamefont{Casati}},
  \emph{\bibinfo{title}{Thermal diode: rectification of heat flux}},
  \bibinfo{journal}{Phys. Rev. Lett.} \textbf{\bibinfo{volume}{93}},
  \bibinfo{pages}{184301} (\bibinfo{year}{2004}).

\bibitem[{\citenamefont{Segal}(2006)}]{seg06}
\bibinfo{author}{\bibfnamefont{D.}~\bibnamefont{Segal}},
  \emph{\bibinfo{title}{Heat flow in nonlinear molecular junctions: Master
  equation analysis}}, \bibinfo{journal}{Phys. Rev. B}
  \textbf{\bibinfo{volume}{73}}, \bibinfo{pages}{205415}
  (\bibinfo{year}{2006}).

\bibitem[{\citenamefont{Ruokola et~al.}(2009)\citenamefont{Ruokola, Ojanen, and
  Jauho}}]{ruo09}
\bibinfo{author}{\bibfnamefont{T.}~\bibnamefont{Ruokola}},
  \bibinfo{author}{\bibfnamefont{T.}~\bibnamefont{Ojanen}}, \bibnamefont{and}
  \bibinfo{author}{\bibfnamefont{A.-P.} \bibnamefont{Jauho}},
  \emph{\bibinfo{title}{Thermal rectification in nonlinear quantum circuits}},
  \bibinfo{journal}{Phys. Rev. B} \textbf{\bibinfo{volume}{79}},
  \bibinfo{pages}{144306} (\bibinfo{year}{2009}).

\bibitem[{\citenamefont{Kuo and Chang}(2010)}]{kuo10}
\bibinfo{author}{\bibfnamefont{D.~M.-T.} \bibnamefont{Kuo}} \bibnamefont{and}
  \bibinfo{author}{\bibfnamefont{Y.-C.} \bibnamefont{Chang}},
  \emph{\bibinfo{title}{Thermoelectric and thermal rectification properties of
  quantum dot junctions}}, \bibinfo{journal}{Phys. Rev. B}
  \textbf{\bibinfo{volume}{81}}, \bibinfo{pages}{205321}
  (\bibinfo{year}{2010}).

\bibitem[{\citenamefont{Whitney}(2013)}]{whi13}
\bibinfo{author}{\bibfnamefont{R.~S.} \bibnamefont{Whitney}},
  \emph{\bibinfo{title}{Nonlinear thermoelectricity in point contacts at pinch
  off: A catastrophe aids cooling}}, \bibinfo{journal}{Phys. Rev. B}
  \textbf{\bibinfo{volume}{88}}, \bibinfo{pages}{064302}
  (\bibinfo{year}{2013}).

\bibitem[{\citenamefont{L{\'o}pez and S{\'a}nchez}(2013)}]{lop13}
\bibinfo{author}{\bibfnamefont{R.}~\bibnamefont{L{\'o}pez}} \bibnamefont{and}
  \bibinfo{author}{\bibfnamefont{D.}~\bibnamefont{S{\'a}nchez}},
  \emph{\bibinfo{title}{Nonlinear heat transport in mesoscopic conductors:
  {Rectification, Peltier effect, and Wiedemann-Franz} law}},
  \bibinfo{journal}{Phys. Rev. B} \textbf{\bibinfo{volume}{88}},
  \bibinfo{pages}{045129} (\bibinfo{year}{2013}).

\bibitem[{\citenamefont{Hwang et~al.}(2013)\citenamefont{Hwang, S{\'a}nchez,
  Lee, and L{\'o}pez}}]{hwa13}
\bibinfo{author}{\bibfnamefont{S.-Y.} \bibnamefont{Hwang}},
  \bibinfo{author}{\bibfnamefont{D.}~\bibnamefont{S{\'a}nchez}},
  \bibinfo{author}{\bibfnamefont{M.}~\bibnamefont{Lee}}, \bibnamefont{and}
  \bibinfo{author}{\bibfnamefont{R.}~\bibnamefont{L{\'o}pez}},
  \emph{\bibinfo{title}{Magnetic-field asymmetry of nonlinear thermoelectric
  and heat transport}}, \bibinfo{journal}{New. J. Phys.}
  \textbf{\bibinfo{volume}{15}}, \bibinfo{pages}{105012}
  (\bibinfo{year}{2013}).

\bibitem[{\citenamefont{Whitney}(2014)}]{whi14}
\bibinfo{author}{\bibfnamefont{R.~S.} \bibnamefont{Whitney}},
  \emph{\bibinfo{title}{Most efficient quantum thermoelectric at finite power
  output}}, \bibinfo{journal}{Phys. Rev. Lett.} \textbf{\bibinfo{volume}{112}},
  \bibinfo{pages}{130601} (\bibinfo{year}{2014}).

\bibitem[{\citenamefont{Utsumi et~al.}(2014)\citenamefont{Utsumi,
  Entin-Wohlman, Aharony, Kubo, and Tokura}}]{uts14}
\bibinfo{author}{\bibfnamefont{Y.}~\bibnamefont{Utsumi}},
  \bibinfo{author}{\bibfnamefont{O.}~\bibnamefont{Entin-Wohlman}},
  \bibinfo{author}{\bibfnamefont{A.}~\bibnamefont{Aharony}},
  \bibinfo{author}{\bibfnamefont{T.}~\bibnamefont{Kubo}}, \bibnamefont{and}
  \bibinfo{author}{\bibfnamefont{Y.}~\bibnamefont{Tokura}},
  \emph{\bibinfo{title}{Fluctuation theorem for heat transport probed by a
  thermal probe electrode}}, \bibinfo{journal}{Phys. Rev. B}
  \textbf{\bibinfo{volume}{89}}, \bibinfo{pages}{205314}
  (\bibinfo{year}{2014}).

\bibitem[{\citenamefont{Werlang et~al.}(2014)\citenamefont{Werlang, Marchiori,
  Cornelio, and Valente}}]{wer14}
\bibinfo{author}{\bibfnamefont{T.}~\bibnamefont{Werlang}},
  \bibinfo{author}{\bibfnamefont{M.~A.} \bibnamefont{Marchiori}},
  \bibinfo{author}{\bibfnamefont{M.~F.} \bibnamefont{Cornelio}},
  \bibnamefont{and} \bibinfo{author}{\bibfnamefont{D.}~\bibnamefont{Valente}},
  \emph{\bibinfo{title}{Optimal rectification in the ultrastrong coupling
  regime}}, \bibinfo{journal}{Phys. Rev. E} \textbf{\bibinfo{volume}{89}},
  \bibinfo{pages}{062109} (\bibinfo{year}{2014}).

\bibitem[{\citenamefont{Eich et~al.}(2014)\citenamefont{Eich, Principi,
  Di~Ventra, and Vignale}}]{eic14}
\bibinfo{author}{\bibfnamefont{F.~G.} \bibnamefont{Eich}},
  \bibinfo{author}{\bibfnamefont{A.}~\bibnamefont{Principi}},
  \bibinfo{author}{\bibfnamefont{M.}~\bibnamefont{Di~Ventra}},
  \bibnamefont{and} \bibinfo{author}{\bibfnamefont{G.}~\bibnamefont{Vignale}},
  \emph{\bibinfo{title}{Luttinger-field approach to thermoelectric transport in
  nanoscale conductors}}, \bibinfo{journal}{Phys. Rev. B}
  \textbf{\bibinfo{volume}{90}}, \bibinfo{pages}{115116}
  (\bibinfo{year}{2014}).

\bibitem[{\citenamefont{Gergs et~al.}(2014)\citenamefont{Gergs, Hörig,
  Wegewijs, and Schuricht}}]{ger14}
\bibinfo{author}{\bibfnamefont{N.~M.} \bibnamefont{Gergs}},
  \bibinfo{author}{\bibfnamefont{C.~B.~M.} \bibnamefont{Hörig}},
  \bibinfo{author}{\bibfnamefont{M.~R.} \bibnamefont{Wegewijs}},
  \bibnamefont{and}
  \bibinfo{author}{\bibfnamefont{D.}~\bibnamefont{Schuricht}},
  \emph{\bibinfo{title}{Charge fluctuations in nonlinear heat transport}},
  \bibinfo{journal}{arXiv:1407.8284 (preprint)}  (\bibinfo{year}{2014}).

\bibitem[{\citenamefont{Biele et~al.}(2014)\citenamefont{Biele, D'Agosta, and
  Rubio}}]{bie14}
\bibinfo{author}{\bibfnamefont{R.}~\bibnamefont{Biele}},
  \bibinfo{author}{\bibfnamefont{R.}~\bibnamefont{D'Agosta}}, \bibnamefont{and}
  \bibinfo{author}{\bibfnamefont{A.}~\bibnamefont{Rubio}},
  \emph{\bibinfo{title}{Time-dependent thermal transport theory}},
  \bibinfo{journal}{arXiv:1412.5765 (preprint)}  (\bibinfo{year}{2014}).

\bibitem[{\citenamefont{Cuevas and Scheer}(2010)}]{cue10}
\bibinfo{author}{\bibfnamefont{J.~C.} \bibnamefont{Cuevas}} \bibnamefont{and}
  \bibinfo{author}{\bibfnamefont{E.}~\bibnamefont{Scheer}},
  \emph{\bibinfo{title}{Molecular electronics: an introduction to theory and
  experiment}} (\bibinfo{publisher}{World Scientific}, \bibinfo{year}{2010}).

\bibitem[{\citenamefont{Galperin et~al.}(2009)\citenamefont{Galperin, Saito,
  Balatsky, and Nitzan}}]{gal09}
\bibinfo{author}{\bibfnamefont{M.}~\bibnamefont{Galperin}},
  \bibinfo{author}{\bibfnamefont{K.}~\bibnamefont{Saito}},
  \bibinfo{author}{\bibfnamefont{A.~V.} \bibnamefont{Balatsky}},
  \bibnamefont{and} \bibinfo{author}{\bibfnamefont{A.}~\bibnamefont{Nitzan}},
  \emph{\bibinfo{title}{Cooling mechanisms in molecular conduction junctions}},
  \bibinfo{journal}{Phys. Rev. B} \textbf{\bibinfo{volume}{80}},
  \bibinfo{pages}{115427} (\bibinfo{year}{2009}).

\bibitem[{\citenamefont{Matthews et~al.}(2014)\citenamefont{Matthews, Battista,
  S{\'a}nchez, Samuelsson, and Linke}}]{matthews}
\bibinfo{author}{\bibfnamefont{J.}~\bibnamefont{Matthews}},
  \bibinfo{author}{\bibfnamefont{F.}~\bibnamefont{Battista}},
  \bibinfo{author}{\bibfnamefont{D.}~\bibnamefont{S{\'a}nchez}},
  \bibinfo{author}{\bibfnamefont{P.}~\bibnamefont{Samuelsson}},
  \bibnamefont{and} \bibinfo{author}{\bibfnamefont{H.}~\bibnamefont{Linke}},
  \emph{\bibinfo{title}{Experimental verification of reciprocity relations in
  quantum thermoelectric transport}}, \bibinfo{journal}{Phys. Rev. B}
  \textbf{\bibinfo{volume}{90}}, \bibinfo{pages}{165428}
  (\bibinfo{year}{2014}).

\bibitem[{\citenamefont{Sothmann et~al.}(2015)\citenamefont{Sothmann,
  S{\'a}nchez, and Jordan}}]{sot14}
\bibinfo{author}{\bibfnamefont{B.}~\bibnamefont{Sothmann}},
  \bibinfo{author}{\bibfnamefont{R.}~\bibnamefont{S{\'a}nchez}},
  \bibnamefont{and} \bibinfo{author}{\bibfnamefont{A.~N.}
  \bibnamefont{Jordan}}, \emph{\bibinfo{title}{Thermoelectric energy harvesting
  with quantum dots}}, \bibinfo{journal}{Nanotechnology}
  \textbf{\bibinfo{volume}{26}}, \bibinfo{pages}{032001}
  (\bibinfo{year}{2015}).

\end{thebibliography}
\bibliographystyle{apsrevTitle}

\end{document}